\documentclass[%
superscriptaddress,
 amsmath,amssymb,
 aps,
prb,
preprint,
]{revtex4-2}

\usepackage{graphicx}
\usepackage{dcolumn}
\usepackage{bm}

\begin{document}


\title{Impact ionization in low-band-gap semiconductors driven by ultrafast THz excitation: beyond the ballistic regime}

\author{Simone Biasco}
\affiliation{Institute for Quantum Electronics, Physics Department, ETH Zurich, CH-8093 Zurich, Switzerland.}
\author{Florence Burri}
\affiliation{Institute for Quantum Electronics, Physics Department, ETH Zurich, CH-8093 Zurich, Switzerland.}
\author{Sarah Houver}
\affiliation{Institute for Quantum Electronics, Physics Department, ETH Zurich, CH-8093 Zurich, Switzerland.}
\affiliation{Université Paris Cité, CNRS, Matériaux et Phénomènes Quantiques, F-75013, Paris, France}
\author{Elsa Abreu}
\affiliation{Institute for Quantum Electronics, Physics Department, ETH Zurich, CH-8093 Zurich, Switzerland.}
\author{Matteo Savoini}
\affiliation{Institute for Quantum Electronics, Physics Department, ETH Zurich, CH-8093 Zurich, Switzerland.}
\author{Steven L. Johnson}
\email{johnson@phys.ethz.ch}
\affiliation{Institute for Quantum Electronics, Physics Department, ETH Zurich, CH-8093 Zurich, Switzerland.}
\affiliation{SwissFEL, Paul Scherrer Institut, CH-5232 Villigen PSI, Switzerland}

\date{\today}

\begin{abstract}
Using two-dimensional THz spectroscopy in combination with numerical models, we investigate the 
dynamics linked to carrier multiplication caused by high-field THz excitation of the 
low-gap semiconductor InSb.  In addition to previously observed dynamics connected with quasi-ballistic 
carrier dynamics, we observe other spectral and temporal features that we attribute to impact ionization for peak fields above 
60~kV/cm, which continue up to the maximum investigated peak field of 430~kV/cm.  At the highest fields 
we estimate a carrier multiplication factor greater than 10 due to impact ionization, which is well-reproduced by a numerical simulation of the impact ionization process which we have developed. 
\end{abstract}

\maketitle


\section{\label{sec:intro}Introduction}

The development of new radiation sources and detectors in the terahertz (THz) frequency range, a 
previously under-explored region of the electromagnetic spectrum between 0.3~THz and 30~THz~\cite{10.1038/nphoton.2007.3,10.1088/0034-4885/69/2/r01}, has 
paved the way for new possibilities to investigate solid-state systems on timescales of a few 
picoseconds. The development of intense single-cycle THz sources and their use in ultrafast terahertz 
time-domain spectroscopy (THz TDS) have offered a powerful approach to study the properties of 
semiconductors in different physical regimes relevant to a number of scientific and technological 
fields. Intense THz electric fields have been used to excite and investigate different transport 
effects in semiconductors, for example ballistic electron acceleration in GaAs and InGaAs~\cite{10.1103/PhysRevLett.104.146602,10.1103/PhysRevLett.107.107401}, 
interband tunneling in GaAs~\cite{10.1103/PhysRevB.82.075204}, as well as impact ionization and intervalley scattering in 
semiconductors with direct and indirect energy bandgaps~\cite{10.1103/PhysRevB.81.035201,10.1109/JPHOT.2010.2050873}, including InSb~\cite{10.1103/PhysRevB.78.125203,10.1103/PhysRevB.79.161201}.

In recent years, two-dimensional terahertz time-domain spectroscopy (2D THz TDS) has emerged as a 
variant of the basic TDS technique which is able to study the nonlinear electronic dynamics of 
crystalline materials~\cite{10.1088/1367-2630/15/2/025039}. It has been applied to the study of diverse phenomena such as electron-
phonon coupling in GaAs/AlGaAs quantum wells~\cite{10.1103/PhysRevLett.107.067401}, two-phonon coherence in InSb~\cite{10.1103/PhysRevLett.116.177401}. and soft modes in 
ferroelectrics~\cite{10.1103/PhysRevX.11.021023}.

Here we consider applications of 2D THz TDS to the study of nonlinear carrier dynamics in InSb. 
InSb is typical of a broad class of III-V low-bandgap semiconductors, and has become 
central to a number of technological applications, such as fast photodetectors~\cite{10.1016/j.infrared.2018.12.034,10.1063/1.2737768} and high-speed 
electronic devices with low-power consumption~\cite{10.1063/1.114063,10.1049/el:20071335,10.1049/el:20071335,10.1063/1.3402760}. The material is characterized by a strong spin-orbit 
coupling, large bulk electron mobility~\cite{10.1021/acs.nanolett.5b05125,10.1103/PhysRevB.77.033204}, and a very small electron effective mass~\cite{10.1063/1.95789}. Due to 
these peculiar physical properties, InSb has become a prototypical material for the study of 
fundamental transport and optical effects in narrow-bandgap semiconductors, and a promising platform 
for novel applications in a wide range of fields, such as spintronics~\cite{10.1103/RevModPhys.76.323}, magneto-plasmonics~\cite{10.1063/1.4968178} and 
superconducting devices~\cite{10.1038/s41467-019-11742-4,10.1063/5.0071218}. 

Recently, 2D THz TDS has been used to investigate the low-field ballistic electronic transport regime 
in InSb and its coupling to the lattice excitations, reconstructing the local curvature of the 
conduction band in detail~\cite{10.1364/OE.27.010854}. In this regime (peak fields below 60~kV/cm) the dynamics are accurately 
modeled by assuming the electrons are driven coherently and ballistically by the oscillating THz 
fields.  
In this work, we report a new investigation of the ultrafast nonlinear response of bulk InSb at room 
temperature with 2D THz TDS experiments with THz pump peak fields ranging between 60~kV/cm and 
430~kV/cm. In this regime 2D THz spectra indicate the onset of carrier multiplication processes such as 
impact ionization and interband tunneling.  We observe a significant broadening and shift of the plasma 
resonance, which we argue can be identified as a spectral fingerprint of the THz-driven increase of the 
conduction band population becoming up to 70 times larger than its equilibrium value. The experimental 
nonlinear traces are analyzed in detail in both time and frequency domains, highlighting the relevant 
timescales of carrier multiplication and its effect on the material reflectivity. 

We compare these data with the results of a numerical simulation where we solve Maxwell’s equations 
taking into account the effects of both impact ionization and ballistic carrier transport. This hybrid 
model predicts the increase of the intensity and the broadening of the plasma spectral peak for high 
THz pump levels, in very good agreement with the measured 2D nonlinear spectra.  Moreover, the 
calculated increase in the electron population is compatible with the values retrieved from the 
experimental data, indicating that the hybrid simulation gives a reasonable description of the material 
response.

\section{\label{sec:expmethods}Experimental Methods}

Our 2D THz spectroscopy setup employs two independent THz pump and THz probe pulses generated from an 
amplified femtosecond laser~\cite{10.1364/OE.27.010854}. A Ti:Sapphire amplifier operating at 800~nm with repetition rate of 1 
kHz and 100-fs pulse duration is used to seed two optical parametric amplifiers (OPAs). The signal 
outputs of the two OPAs are independently tuned to emit at wavelengths of 1.3~\(\mu\)m and 1.5~\(\mu\)m 
with energies of 0.8~mJ/pulse and 1.2~mJ/pulse, respectively.

We use the 1.3~\(\mu\)m-beam to generate broadband THz pulses using the ``air-plasma'' method~\cite{10.1364/OL.29.001120}.  
This involves focusing the beam with an off-axis parabolic mirror of focal length \(f = 8\) cm through 
a BBO crystal that is type-I phase matched for second-harmonic generation to a focal point in ambient 
air.  The resulting two-color interaction with the air plasma results in a broadband THz pulse with a 
typical peak field intensity up to 100 kV/cm and broadband spectrum extending over the range 1-12 THz. 
The 1.5 \(\mu\)m-beam is used to pump a DSTMS organic crystal, where optical rectification (OR) induces 
single-cycle THz pulse generation~\cite{10.1002/adfm.200601117} with peak intensities up to 450 kV/cm and a relatively narrow 
bandwidth covering the frequency interval between 1.5 and 4.5 THz. The polarization of the air-plasma 
THz field is horizontal, whereas the OR-generated THz field is vertically polarized. In our 2D THz THD system the narrow-band OR-generated THz field is used as a high-intensity pump to 
excite the sample out of equilibrium, while the broadband air-plasma-generated THz pulse probes the 
broadband response of the system, with a focused peak field amplitude typically 2 to 10 times weaker 
than that of the pump. 

The two cross-polarized THz beams are brought onto the sample using a combining wire-grid polarizer 
(WG), that reflects the THz pump  and transmits the THz probe beams. After both beams reflect from the 
sample, the same WG lets the reflected probe beam through, while filtering out the reflection of the 
vertically polarized THz pump beam. The reflected probe signal is then refocused into an Air-Biased 
Coherent Detection (ABCD) system, which measures a projection of the time-dependent electric field at 
the center of the reflected beam~\cite{10.1103/PhysRevLett.97.103903}. This broadband, phase-resolved detection scheme uses an infrared 
detection beam, obtained by sampling a portion of the 1.3 \(\mu\)m-beam with a 90-10 beamsplitter, to 
characterize the reflected THz pulse from the sample. 

The ABCD data acquisition architecture is based on a hardware solution that chops the THz probe beam at
500 Hz and the THz pump beam at 250 Hz with fixed phase relationship, while the polarity of the ABCD 
bias is switched at a frequency of 125 Hz~\cite{10.1364/OE.27.010854}. The choppers alternately expose the sample to different 
combinations of THz-light exposure (plasma-generated pulse only, OR pulse only, both pulses, and no 
pulses).  The ABCD setup measures a selected projection of the time-dependent electric field at the 
center of the reflected beams from each of these combinations: the plasma-generated probe pulse 
\(E_\textrm{probe}^\prime (t)\), the OR-generated pump \(E_\textrm{pump}^\prime (t)\), the 
superposition of both temporal signals \(E_\textrm{pump+probe}^\prime (t)\), as well as the background 
signal with all fields blocked. The projection axis of the ABCD is aligned to the polarization of the 
plasma-generated THz field. The pump-probe delay \(\tau\) between the two independent THz beams and the 
detection delay t between the THz pulses and the infrared detection beam can be independently 
controlled.  Here we define \(\tau\) as the time difference between the arrival of the maximum field 
of the two pulses at the sample, with positive values indicating that the pump arrives first.  The time 
\(t = 0\) is similarly defined as the time when the ABCD sampling pulse is coincident with the maximum 
of the probe pulse. For our 2D measurement in reflection mode, we define the “nonlinear” component of the field response as
\begin{equation}
E_\textrm{NL}^\prime(t,\tau) = E_\textrm{pump+probe}^\prime(t,\tau) - E_\textrm{pump}^\prime (t,\tau) - E_\textrm{probe}^\prime(t,\tau)\label{eq:NLprime}.
\end{equation}

To calibrate the absolute THz peak field amplitudes at the sample position we use electro-optic 
sampling in a 100 \(\mu\)m-thick GaP crystal placed at the sample position. In order to investigate the 
high-field effects in the semiconductor sample, we control the incident THz pump intensity with a pair 
of WG polarizer on the path of the OR-generated beam before the sample. The first WG is set at an angle 
\(\theta\) with respect to the vertical, whereas the second polarizer is set to pass vertically 
polarized light. The transmitted field is then nominally attenuated to \(\cos^2\theta\) of the OR-generated field, with the actual field amplitude directly measured by electro-optic sampling at the 
sample location. In this way THz pump peak field amplitudes are varied in the range between 60~kV/cm 
and 430~kV/cm, while keeping the THz probe fixed to a lower peak field value of 40~kV/cm.

In order to optimize the signal-to-noise ratio of the measurements, the TDS setup is enclosed in a 
nitrogen-purged box encompassing the THz beam paths, so that the relative humidity is kept below 1.8\% 
and absorption due to water molecules and other THz-active chemicals is strongly reduced. 

\section{\label{sec:Results}Results}

Using our 2D THz TDS setup, we investigate the nonlinear signal reflected from a bulk InSb sample, 
nominally undoped and kept at room temperature. The 350 \(\mu\)m-thick sample is oriented with its
(100)-cut surface normal to both incident THz pulses. The polarization of the probe air-plasma pulse is 
along the [010] direction.  The optical-rectification-generated pump is oriented along [001]. 

\begin{figure}
    \centering
    \includegraphics[width=14cm]{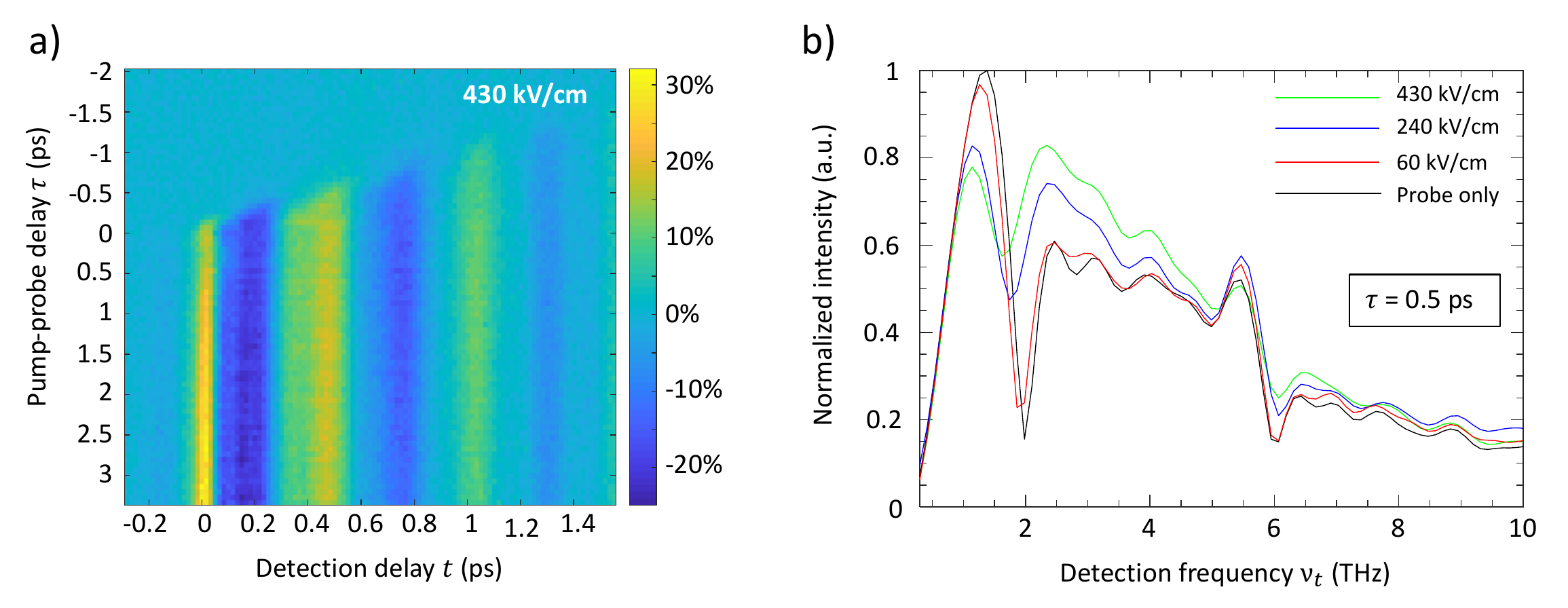}
    \caption{2D time traces and 1D nonlinear probe spectra.
a) Nonlinear signal \(E_\textrm{NL}^\prime(t,\tau)\) reflected from InSb measured with cross-
polarized THz probe \(E_\textrm{probe}\)   and THz pump \(E_\textrm{pump}\) fields, along (010) 
and (001), respectively. The nonlinear 2D temporal trace is reported as a function of detection 
delay \(t\) and pump-probe delay \(\tau\), with a colormap renormalized to the maximum of the 2D 
temporal trace of \(E_\textrm{probe}\). The nonlinear signal is measured for peak field amplitudes \(E_\textrm{probe, max} = 40\) kV/cm and \(E_\textrm{pump, max} = 430\)~kV/cm  in a nitrogen-purged 
atmosphere. b) Plot of the probe spectrum alone (black) and the nonlinear probe spectrum at 
fixed pump-probe delay \(\tau = 0.5\)~ps for THz pump amplitudes of 60~kV/cm (red), 240~kV/cm 
(blue) and 430~kV/cm (green). }
    \label{fig:2Dand1D}
\end{figure}

Fig.~\ref{fig:2Dand1D}a reports the 2D temporal trace of the reflected nonlinear signal \(E_\textrm{NL}^\prime\) 
for a pump peak 
field of 430~kV/cm and a probe peak field of 40 kV/cm. The nonlinear response is measured for pump-probe 
delays \(\tau\) in the range between -2 ps and 3.35 ps, and for detection delays \(t\) between \(-0.3\)~ps and \(1.55\)~ps, 
with time-steps of \(\Delta \tau = 66\)~fs and \(\Delta t  = 16\)~fs. The 2D trace shows a strong modulation of the nonlinear 
field along the detection time axis, with alternating sign and a maximum modulation up to \(\pm 30\)\% of the 
plasma field maximum amplitude. Sharp features at \(t=0\)~ps, 0.22~ps and 0.45~ps indicate that the high 
pump intensity excites a fast temporal response from the material, which in turn corresponds to a 
stronger spectral density of the high-frequency components of \(E_\textrm{NL}^\prime\). 
This enhancement of the higher 
frequency components in the material response increases with the applied THz pump peak field, and can be 
quantified by examining the spectra after the excitation pulse at a fixed pump-probe delay of 
\(\tau = 0.5\)~ps. To show more clearly this effect we define the ``nonlinear probe signal''
\begin{equation}
E_\textrm{NL probe}^\prime (t,\tau) = E_\textrm{pump+probe}^\prime(t,\tau) - E_\textrm{pump}^\prime(t,\tau)\label{eq:NLprobe}.
\end{equation}

Fig.~\ref{fig:2Dand1D}b shows the one-dimensional Fourier transform of \(E_\textrm{NL probe}^\prime(t,\tau)\) with 
respect to detection time \(\tau\) for a fixed pump-probe delay \(\tau = 0.5\)~ps in comparison to the 
corresponding Fourier transform of the reflected plasma-generated electric field alone 
\(E_\textrm{probe}^\prime(t,\tau=0.5\textrm{ ps})\), highlighting 
the difference between the excited material response and its unpumped response. These comparisons are 
shown for a THz probe peak field of 40 kV/cm and THz pump peak fields of 60~kV/cm, 240~kV/cm and 
430~kV/cm. One general trend is a pump-field dependent red-shift of the plasma edge at 
2~THz, 
which manifests in the spectrum as a sharp minimum in the spectral density.  At the highest pump peak 
field this red-shift is approximately 20\%.  This is accompanied by a broadband increase in the spectral 
density between 2.5 and 6~THz. Indeed, the increase of the high-frequency components for high THz pump 
amplitudes extends even beyond the LO phonon resonance at 5.7~THz, which also undergoes an intensity 
attenuation with increasing pump field.

\begin{figure}
    \centering
    \includegraphics[width=14cm]{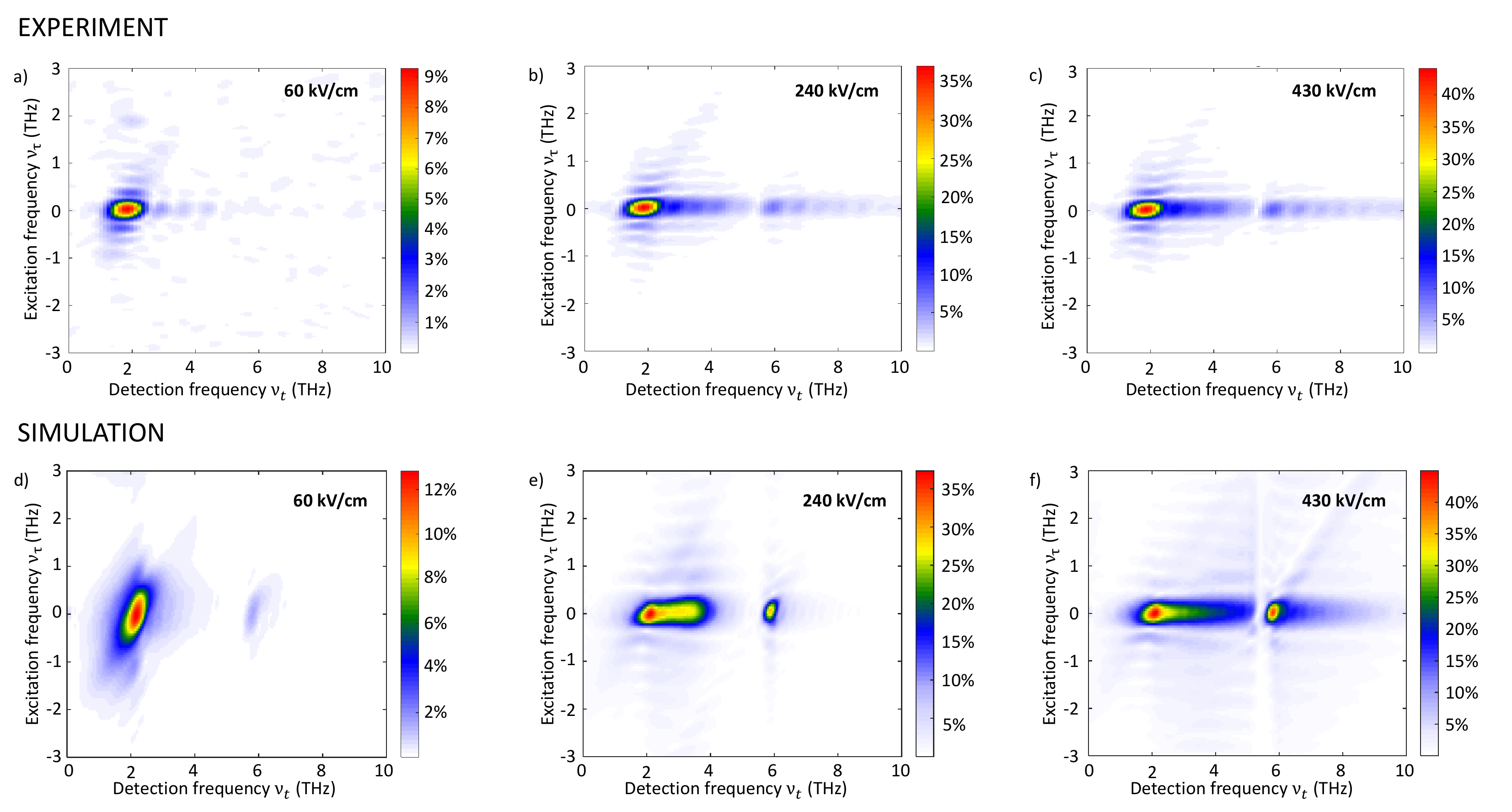}
    \caption{Simulated and experimental 2D spectra
a) Experimental 2D nonlinear spectrum obtained from the Fourier transform of the experimental 
\(E_\textrm{NL}^\prime(t,\tau)\) as a function of detection frequency \(\nu_t\) and excitation frequency 
\(\nu_\tau\). The color scale is normalized to the maximum of the 2D spectrum of the probe 
field \(E_\textrm{probe}\). This 2D spectrum is obtained with peak field intensities \(E_\textrm{probe, max} = 40\)~kV/cm and \(E_\textrm{pump, max} = 60\)~kV/cm, with the same 
reflection configuration for cross-polarized THz pump and probe. b) Experimental 2D nonlinear 
spectrum  retrieved for 
\(E_\textrm{probe, max}  = 40\)~kV/cm and \(E_\textrm{pump, max} = 240\)~kV/cm. c) Experimental 
2D nonlinear spectrum retrieved for \(E_\textrm{probe, max} = 40\)~kV/cm and \(E_\textrm{pump, max} = 430\)~kV/cm. 
d) Simulated 2D nonlinear spectrum  obtained with cross-polarized THz probe \(E_\textrm{probe, max} =40\)~kV/cm and THz pump \(E_\textrm{pump, max} = 60\)~kV/cm, implemented using a hybrid quasi-ballistic finite-difference time domain (FDTD) 
algorithm that includes equations for the impact ionization mechanism in the semiconductor InSb. 
The color values are normalizes to the maximum of the 2D spectrum of the probe field \(E_\textrm{probe}\). 
e) Simulation of the 2D nonlinear spectrum calculated with 
\(E_\textrm{probe, max} = 40\)~kV/cm and \(E_\textrm{pump, max} = 240\)~kV/cm. 
f) Computed 2D nonlinear spectrum for
\(E_\textrm{probe, max} = 40\)~kV/cm and highest THz pump field 
\(E_\textrm{pump, max} = 430\)~kV/cm, as used in the experiment.
}
    \label{fig:Exp_and_sim}
\end{figure}

The full 2D TDS traces of the nonlinear signal can be analyzed by performing a direct two-dimensional 
Fourier transform, so that the modulations of \(E_\textrm{NL}^\prime (t,\tau)\) as functions of the 
pump-probe and 
detection delays are mapped in their distinct peaks in 2D spectra \(E_\textrm{NL}^\prime (\nu_t,\nu_\tau)\). 
Fig.~\ref{fig:Exp_and_sim}a shows such a two-dimensional spectrum of the nonlinear signal, in the case of lowest THz pump 
peak field intensity (60~kV/cm). The intensity scale is normalized such that a value of 100\% 
corresponds to the maximum value of the amplitude of the 2D Fourier transform of \(E_\textrm{probe}^\prime(t,\tau)\). In 
this regime, the spectrum reveals a well-defined peak at \((\nu_t,\nu_\tau) = (2\textrm{ THz}, 0 \textrm{ THz})\) with an amplitude 
of 10\%. Figs.~\ref{fig:Exp_and_sim}b and \ref{fig:Exp_and_sim}c report the 2D spectra for THz pump peak fields of 240~kV/cm and 430~kV/cm, 
where the plasma peak intensity increases to 40\% and 50\%, respectively. Besides this, an enhancement 
of the nonlinear spectrum along the horizontal line \(\nu_\tau = 0\)~THz is visible in the range between 
\(\nu_t= 2\)~THz and \(\nu_t= 10\)~THz, with a peak 
clearly appearing at \((\nu_t, \nu_\tau) = (6\textrm{ THz}, 0 \textrm{ THz})\) with a relative intensity up 
to 11\% for the maximum THz pump peak of 430~kV/cm.  For this peak the value of \(\nu_t\) is very close to that of the LO phonon frequency (5.7 THz).

Qualitatively, the 2D nonlinear signal shows that increasing levels of the THz pump induce a larger broadband modification of the reflectivity spectrum, with an enhanced spectral density immediately above the plasma edge that extends towards the high-frequency sensitivity limit of our ABCD system. 

\section{\label{sec:Simulation} Simulation}

In order to further interpret the experimental results retrieved with these 2D THz TDS experiments, we 
developed a finite difference time-domain (FDTD) model of the interaction of the THz pump and probe 
pulses with the bulk InSb sample to solve Maxwell’s equations in the system~\cite{10.1109/TAP.1966.1138693,10.1103/PhysRevB.95.125201}. We consider the two 
cross-polarized THz pulses propagating along the z-direction through the thick sample, with the 
electronic response of InSb described by a two-dimensional electronic band structure in the x-y plane. 
For low THz pump fields, the electronic response can be studied in the framework of ballistic motion, 
with a coherent wave packet driven by the applied THz electric field~\cite{10.1364/OE.27.010854}. In this limit, the equation 
of motion for the carriers driven by the THz pulse is solved for the in-plane components of the electron 
wavevector \((k_x,k_y)\), and the corresponding electron energy is retrieved from the band structure, 
obtained from a tight-binding model. 

This picture changes significantly when higher THz pump pulses are applied.  In this case, the strong THz 
OR pump pulse accelerates the conduction electrons to kinetic energies larger than the energy bandgap, 
which is \(\epsilon_g = 170\)~meV in room-temperature InSb. While the small energy of THz photons 
cannot induce direct optical transitions via single-photon processes even in such a low-bandgap 
semiconductor, the THz-driven electrons reach kinetic energies large enough to induce nonlinear carrier 
generation and recombination effects~\cite{10.1007/BF00617768,10.1088/0268-1242/28/2/025019}, sometimes associated with inter-band scattering processes 
that can significantly alter the electron concentration and motion.\cite{10.1038/s41598-020-67541-1} Other experimental works have 
demonstrated that THz pulses can activate the impact ionization mechanism at temperatures below 260~K, 
with an electron population increase that strongly depends on the THz pump pulse duration and intensity~\cite{10.1103/PhysRevB.78.125203,10.1103/PhysRevB.79.161201}.

We expect impact ionization to have a dominant role in the electronic response of InSb when the average 
carrier kinetic energy becomes larger than the characteristic threshold energy \(\epsilon_\textrm{th} = \epsilon_g\). In 
this condition, the scattering between accelerated carriers inside the material can promote electrons 
from the valence band to the conduction band. In a single scattering event, a high-energy conduction 
electron loses a part of its kinetic energy to create an electron-hole pair. This process can then 
repeat itself with the new carriers, causing an exponential increase in carrier density as long as the 
transient THz pulse is sufficient to drive the newly created carriers to kinetic energies above \(\epsilon_\textrm{th}\). An 
in-depth statistical analysis of such process can be performed with Monte Carlo simulations~\cite{10.1088/0268-1242/28/2/025019,10.1038/s41598-020-67541-1},
where a large number of single scattering events are simulated to average over all possible combinations 
of final carrier energies and momenta, and over all carriers in the different bands. An integration of 
such method using FDTD to reconstruct the propagation of the THz electromagnetic wave in the material in 
presence of impact ionization would require a time-consuming iteration of Monte Carlo simulations at 
each time step, with the updated conditions of the electron energy and momentum. This would severely 
limit the speed of the otherwise rapid FDTD algorithms and likely prove impractical for our purposes. We 
instead here apply an alternative and efficient FDTD approach based on an approximated treatment of the 
impact ionization effects, allowing a fast computation of the electromagnetic fields propagating in the 
InSb sample, as well as the fields reflected by the material. 

We consider a one-dimensional spatial mesh along the pulse propagation direction \(z\), where our 
material 
thickness \(d\) is divided into discrete segments with length \(\Delta z=0.6\,\mu\)m, surrounded by 
air layers, while the detection time domain \(t\) is discretized in steps of duration 
\(\Delta t = 1.0\)~fs. We define \(z=0\) as the initial air-material interface, and set at a boundary 
condition there a value of the 
electric field \(\mathbf{E}_\textrm{THz} (0,t)=\mathbf{E}_\textrm{pump+probe} (t,\tau)\) the 
superposition of the incident OR-generated and plasma-generated THz fields, separated in time by a fixed pump-
probe delay \(\tau\). The electric field at \(z > 0\) and \(t = 0\) is set to zero. We proceed to 
solve Maxwell’s 
equations in the time domain as described in Ref.~\onlinecite{10.1364/OE.27.010854}, 
setting the initial condition of zero fields in the bulk and using the displacement field
\begin{equation}
\mathbf{D}(z,t) = \epsilon_0 \epsilon_\infty \mathbf{E}_\textrm{THz}(z,t) + \mathbf{P}_\textrm{ph}(z,t) + \mathbf{P}_e(z,t)\label{eq:Displacement}
\end{equation}
where \(\epsilon_0\) is the permittivity of free space, \(\epsilon_\infty = 15.78\) is the background 
dielectric constant as determined from an analysis of independent linear THz TDS measurements~\cite{10.1364/OE.27.010854},
\(\mathbf{E}_\textrm{THz}\) is the driving field, \(\mathbf{P}_\textrm{ph}\) is the polarization density 
due to the lowest-frequency infrared-active vibrational mode while 
\(\mathbf{P}_e\) is the electronic polarization density arising from free carriers. We determine 
\(\mathbf{P}_\textrm{ph} (z,t)\) via the auxiliary partial differential equation
\begin{equation}
\frac{\partial^2\mathbf{P}_\textrm{ph}}{\partial t^2} + \frac{1}{\tau_\textrm{ph}} \frac{\partial \mathbf{P}_\textrm{ph}}{\partial t} + 4 \pi^2\nu_\textrm{TO}^2 \mathbf{P}_\textrm{ph} = 4 \pi^2 \epsilon_\infty (\nu_\textrm{LO}^2 - \nu_\textrm{TO}^2) \mathbf{E}_\textrm{THz}\label{eq:phonon}
\end{equation}
 for the initial condition \(\mathbf{P}_\textrm{ph}(z,0) = 0\).  Here we take 
 \(\nu_\textrm{TO} = 5.34\)~THz, \(\nu_\textrm{LO} = 5.70\)~THz, 
 and \(\tau_\textrm{ph} = 0.26\)~ps, values determined from linear THz TDS measurements~\cite{10.1364/OE.27.010854}.

Next we discuss how to determine the free carrier contribution \(\mathbf{P}_e(z,t)\) to the polarization 
density.  InSb has a 0.17 eV direct band gap at the \(\Gamma\) point~\cite{10.1063/1.95789}.  There is a single lowest
energy conduction band, and two highest-energy valence bands that are degenerate at \(\Gamma\)~\cite{10.1103/PhysRevB.80.035203}.  
One of the valence bands has a much higher hole effective mass (by a factor of about 40) than either 
the conduction band or the other valence band.  Due to this high effective mass, in thermal equilibrium 
the hole population of the heavy-hole band is much higher than that of the light-hole band.  Similarly, 
the main channel for 
impact ionization is the ionization of heavy holes from collisions with highly energetic conduction band 
electrons~\cite{10.1016/0022-3697(57)90013-6}. For these reasons we neglect the light-hole band in our analysis, and identify the main 
contributions to  \(\mathbf{P}_e(z,t)\) to come from the dynamics of electrons in the conduction band that 
occasionally lose energy via ionization of heavy-hole-band electrons.  The very large effective mass of 
the heavy holes relative to the conduction electrons makes it possible to neglect their direct 
contribution to \(\mathbf{P}_e(z,t)\).

To model the dynamics of the conduction electrons and their contribution to the polarization density, we 
employ a simplified model similar to that used in Houver et al.~\cite{10.1364/OE.27.010854} but with modifications to include 
the effects of carrier multiplication due to impact ionization.   Rather than attempt to model a 
statistical distribution of such electrons at a particular depth \(z\) and time \(t\), we instead approximate 
them as a wavepacket consisting of a local density of \(N(z,t)\) electrons all with an identical crystal 
momentum \(\hbar \mathbf{k}(z,t)\).  The energy per electron \(\epsilon(z,t)\) is then determined by the 
conduction band dispersion.  
We assume the electrons obey a classical equation of motion of the form
\begin{equation}
\hbar \frac{\partial \mathbf{k}}{\partial t} + \frac{\hbar}{\tau_\epsilon}\mathbf{k}-\mathbf{F}_\textrm{imp}(\mathbf{k}) = - e \mathbf{E}_\textrm{THz}\label{eq:EOM}
\end{equation}
where \(1/\tau_\epsilon\) is an intrinsic damping rate determined by the electron energy relaxation 
time \(\tau_\epsilon = 5\)~ps in the absence of impact ionization\cite{10.1103/PhysRevB.79.161201}, \(e\) is the elementary charge, 
and \(\mathbf{F}_\textrm{imp}(\mathbf{k})\) is an effective damping force associated with impact 
ionization scattering processes, whose explicit form will be derived below. We set the initial condition 
\(\mathbf{k}(z,0) = 0\), which means the electron wavepacket is initially at the bottom of the 
conduction band at the \(\Gamma\) point. The time-domain polarization density \(\mathbf{P}_e\) is 
then determined by
\begin{equation}
\frac{\partial \mathbf{P}_e}{\partial t} = -e N \mathbf{v}\label{eq:Pe}
\end{equation}
where \(N\) is the density of conduction-band electrons and \(\mathbf{v} = \mathbf{\nabla}_k\epsilon/\hbar\) is the group velocity of the electrons.  As initial conditions we 
set \(\mathbf{P}_e(z,0)=0\). 

We next need to find a way to determine the carrier density N(z,t), which here is explicitly 
time-dependent to allow for carrier multiplication.  Here we will consider only impact ionization as a 
mechanism for carrier multiplication.  Impact ionization is a process that occurs when the kinetic 
energy of a carrier exceeds the minimum energy \(\epsilon_\textrm{th}=0.17\)~eV required to create 
an electron-hole pair, 
resulting in an increase in carrier density and a decrease in the average kinetic energy of the 
carriers.  This leads to an increase in the carrier density \(N(z,t)\) which can be modeled in InSb as~\cite{10.1103/PhysRevB.79.161201,10.1007/BF00617768} 
\begin{equation}
\frac{\partial N}{\partial t} =  (\epsilon-\epsilon_\textrm{th}) \theta(\epsilon-\epsilon_\textrm{th}) C N\label{eq:Neq}
\end{equation}
where \(\theta(\epsilon-\epsilon_\textrm{th})\) is zero when \(\epsilon < \epsilon_\textrm{th}\) and is 
otherwise unity, and \(C=1.8 \times 10^{13}\)~eV\(^2\)s\(^{-1}\) is a material-specific constant computed 
from the matrix elements of the impact ionization cross section for the case of a parabolic band system.\cite{10.1007/BF00617768}  For our simulations we assume an initial value of \(N(z,0)= N_0= 9.2 \times 10^{15}\)~cm\(^{-3}\), as obtained from linear THz TDS measurements~\cite{10.1364/OE.27.010854}. 

Concomitant with the increase in carrier density is a decrease in the average energy of the electrons in the conduction band.  To leading order this leads to an energy loss contribution 
\begin{equation}
\left(\frac{\partial \epsilon}{\partial t}\right)_\textrm{imp} = - \frac{\epsilon-\epsilon_\textrm{th}}{N} \frac{\partial N}{\partial t}\label{eq:impder}
\end{equation}
In order to find a suitable form of \(\mathbf{F}_\textrm{imp}\) in Eq. 5, we approximate the dispersion 
of the conduction band using an analytical formula for the energy dispersion near the \(\Gamma\) 
valley of the conduction band based on the effective mass approximation:
\begin{equation}
(1+\alpha \epsilon) \epsilon = \frac{\hbar^2}{2m^*} \left| \mathbf{k}\right|^2\label{eq:effmass}
\end{equation}
where \(\alpha= 4.1\)~eV\(^{-1}\)~\cite{10.1080/00018737400101371} and the effective mass \(m^* = 0.0159m_e\), and \(m_e\) is the mass of a free electron.  Using this in 
combination with Eq. 8 yields
\begin{equation}
\mathbf{F}_\textrm{imp} = \hbar \left( \frac{\partial \mathbf{k}}{\partial t}\right)_\textrm{imp} = - \frac{\hbar}{2}\frac{(1+2 \alpha \epsilon)(\epsilon-\epsilon_\textrm{th})^3}{(1+\alpha\epsilon)\epsilon}   \theta(\epsilon-\epsilon_\textrm{th}) C \mathbf{k}
\end{equation}
where we also assume that, on average, impact ionization leads to an isotropic redistribution of carrier momentum.

As is apparent from our Eq.~\ref{eq:EOM}, the time evolution of the carrier momentum and the associated 
kinetic energy in our model is controlled by two competing mechanisms: the acceleration of the 
conduction electron wavepacket by the THz electrical field, and the loss of kinetic energy due to 
repeated impact ionization scattering events promoting new electrons in the conduction band. 
Therefore, the carrier energy \(\epsilon\) can repeatedly exceed and then fall below the threshold value \(\epsilon_\textrm{th}\) 
while the THz pump pulse propagates, depending on its peak amplitude and temporal shape, in analogy 
with the approach used in other semiconductors like GaAs~\cite{10.1038/ncomms1598}. Alternative models for impact ionization 
describe the THz-driven change in the carrier energy as a step-like change with magnitude equal to the 
ponderomotive energy~\cite{10.1103/PhysRevB.79.161201} of the THz pump pulse at time \(t=0\), followed by a subsequent monotonic decay  
according to Eq.~\ref{eq:Neq}.  By their nature these ``step-like'' models cannot resolve carrier 
multiplication effects that are extended over the duration of the pulse and are thus not appropriate 
for our experiment. 

After the THz pulse field vanishes and the carrier energy becomes again lower than \(\epsilon_\textrm{th}\), impact 
ionization comes to an end, letting us retrieve the final carrier concentration in all the material 
slices. To estimate the simulated final carrier multiplication, we average the conduction band density 
profile \(N(0,t)\) over the range \(0.4\textrm{ ps} < t < 0.7\textrm{ ps}\) to obtain \(\left<N(0,t)\right>\) and define the carrier multiplication factor 
\begin{equation}
    F_\textrm{sim} = \frac{\left<N(0,t)\right>}{N_0}
\end{equation}
where 
\(N_0\) is the intrinsic free carrier population of the InSb sample at equilibrium.
Considering separately the case when the sample is illuminated by the THz pump only, the THz probe 
only and both of them with a pump-probe delay \(\tau\), we can simulate the system response at the 
detection time \(t\) and retrieve the reflected nonlinear signal \(E_\textrm{NL}^\prime(t,\tau)\). 

Figs.~\ref{fig:Exp_and_sim}d,~\ref{fig:Exp_and_sim}e and~\ref{fig:Exp_and_sim}f illustrate the simulated 2D nonlinear spectra obtained for the different THz 
pump intensities of 60~kV/cm, 240~kV/cm and 430~kV/cm respectively, with a fixed THz probe peak field 
of 40~kV/cm. For the lowest pump field, Fig.~\ref{fig:Exp_and_sim}d shows a strong peak predicted at \((\nu_t, \nu_\tau) = (2\textrm{ THz}, 0 \textrm{ THz})\) for the plasma resonance with relative nonlinear intensity 
of 12\% with respect to the probe spectral maximum, and a much less intense peak at \((\nu_t, \nu_\tau) = (6\textrm{ THz}, 0 \textrm{ THz})\) for the LO-phonon excitation. For the larger THz pump peak 
field of 240~kV/cm shown in Fig.~\ref{fig:Exp_and_sim}e, both these resonances are enhanced, with the plasma peak 
increasing to a relative value of 37\% and showing a broadening to higher frequencies up to \(\nu_t = 4 \textrm{ THz}\), in good qualitative agreement with the experimental data seen in Fig~\ref{fig:Exp_and_sim}b. At the maximum THz pump peak field of 430~kV/cm, Fig.~\ref{fig:Exp_and_sim}f reports a nonlinear signal of about 44\% at the plasma resonance, very close to the experimental value of 43\%, with a significant broadening that overlaps with the phonon spectral peak and extends up to \(\nu_t = 8 \textrm{ THz}\).

\begin{figure}
    \centering
    \includegraphics[width=14cm]{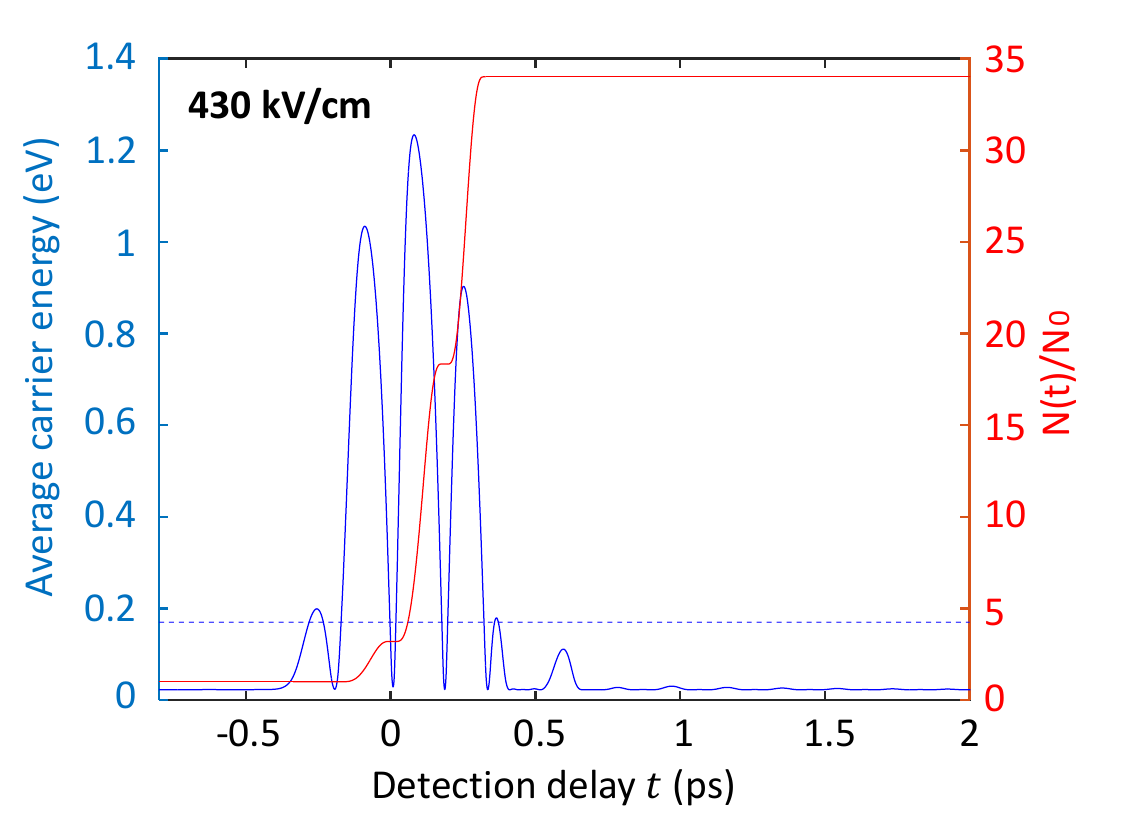}
    \caption{
Plot of the simulated average carrier energy \(\epsilon(t)\) (blue) and the electron concentration 
\(N(t)\) (red) normalized by the initial intrinsic free carrier concentration \(N_0\) as a function of the 
detection time \(t\), for a fixed pump-probe delay \(\tau = 0.5\)~ps. The final multiplication factor of the 
conduction electron population is 34 for \(E_\textrm{probe, max} = 40\)~kV/cm 
and \(E_\textrm{pump, max} = 430\) kV/cm, as impact ionization takes place only over a time window of about 500 fs. 
The horizontal dashed blue line indicates the threshold energy for the impact ionization \(\epsilon_\textrm{th} = 170\)~meV. }
    \label{fig:Sim_Carrier_E}
\end{figure}

An example of the computed evolution of the carrier concentration \(N(t)\) and average carrier energy \(\epsilon(t)\) 
is reported in Fig.~\ref{fig:Sim_Carrier_E} for a pump-probe delay of \(\tau = 0.5 \)~ps, with a THz probe peak field of 40~kV/cm 
and a THz pump peak field of 430~kV/cm, with the electrons initially at the \(\Gamma\) point. Impact ionization 
takes place during discrete time intervals when the carrier kinetic energy is above the 
threshold value, leading to a final multiplication factor of 34. With this intense THz pump 
polarized along the reciprocal axis \(\Gamma \rightarrow \textrm{X}\) of the Brillouin zone, the average kinetic energy reaches a 
maximum of 1.2~eV, larger than the local minima of the L and X valleys, which are 0.68~eV and 1.0~eV 
respectively. While most of the electrons are still confined in the \(\Gamma\) valley, intervalley scattering 
could be possible in this high-pump regime~\cite{10.1038/s41598-020-67541-1,10.1038/ncomms1598},  even if it is not implemented in the model. For an 
intermediate THz pump level of 240~kV/cm, the carrier population grows by approximately 4 times once 
the THz pump pulse has passed and excited them up to a kinetic energy of approximately 0.9~eV. Conversely, the 
lowest THz pump amplitude of 60~kV/cm induces a carrier population increase by  only 1\%, indicating 
a minimal contribution of impact ionization due to very small oscillation of the electron wavepacket 
around the \(\Gamma\) point.

\section{\label{sec:Discussion}Discussion}

\begin{figure}
    \centering
    \includegraphics[width=17cm]{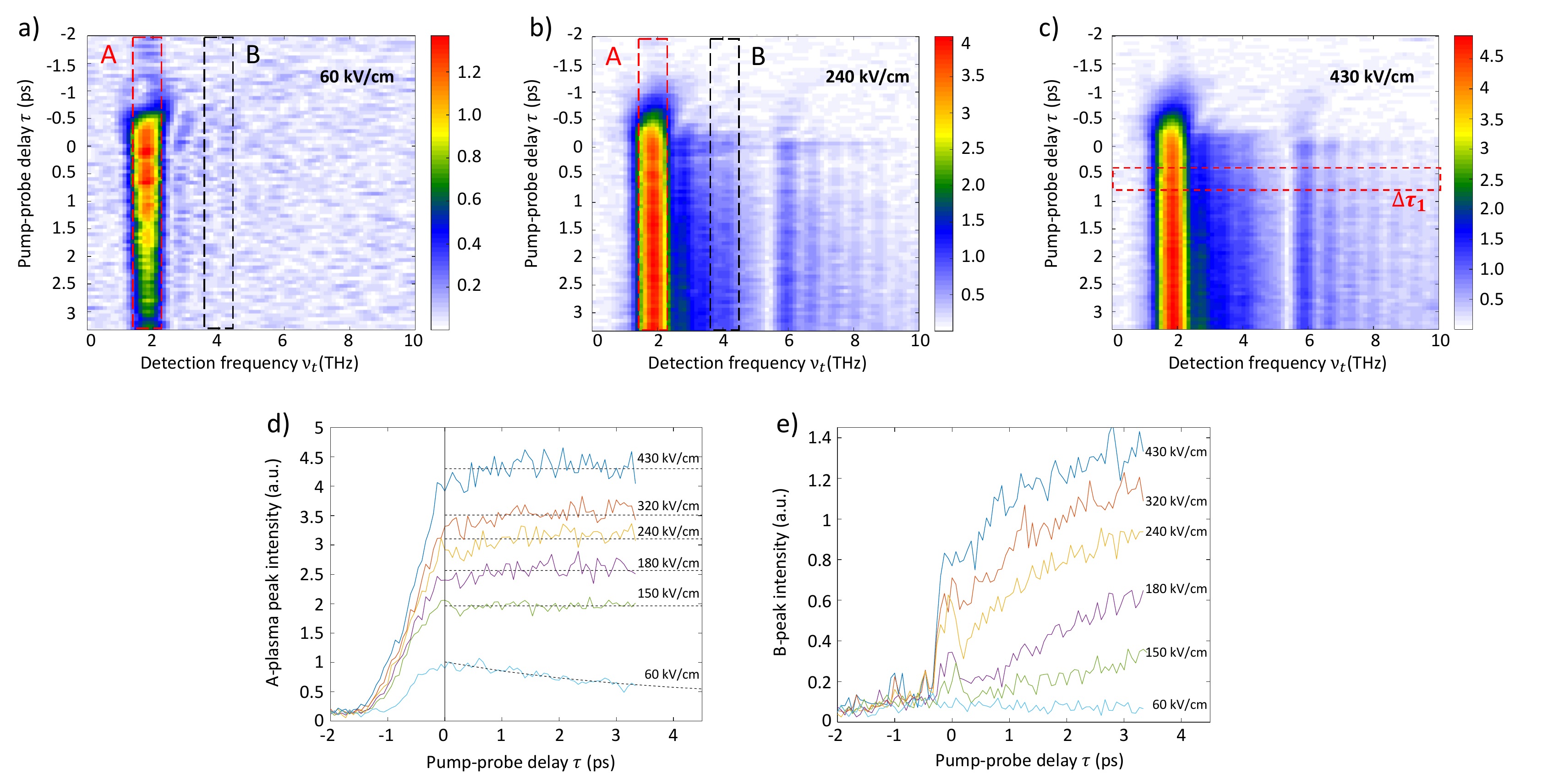}
    \caption{
    a) Spectrogram showing the evolution of the nonlinear spectra along the detection frequency, for different pump-probe delays \(\tau = 0\), for \(E_\textrm{probe, max}   = 40\) kV/cm and \(E_\textrm{pump, max} = 60\) kV/cm. The dashed rectangles identify the detection frequency ranges A (red) and B (black). b) Nonlinear signal spectrogram  retrieved with  \(E_\textrm{probe, max}  = 40\) kV/cm and \(E_\textrm{pump, max} = 240\) kV/cm. c) Nonlinear signal spectrogram retrieved with  \(E_\textrm{probe, max}  = 40\) kV/cm and \(E_\textrm{pump, max} = 430\) kV/cm. d) Evolution of the plasma peak intensity of the spectrograms as a function of \(\tau\) for different THz pump levels, obtained averaging the spectrogram intensity over the detection frequency range A, between \(\nu_t = 1.8\) THz and \(\nu_t = 2.2\) THz. The vertical black line indicate time \(\tau = 0\) ps, while the dashed black lines indicate the fit of the traces with either a decaying exponential function (at 60 kV/cm) or constant lines (for 150 kV/cm and higher). e) Evolution of the intensity of the spectrograms averaged in the detection frequency range B between \(\nu_t = 3.8\) THz and \(\nu_t = 4.2\) THz as a function of \(\tau\) for different THz pump levels in the range 60~kV/cm and 430~kV/cm
}
    \label{fig:spec_and_temp_dec}
\end{figure}

Besides these numerical calculations of the 2D nonlinear spectra, we take advantage of the 2D THz 
TDS scheme to measure the evolution of the nonlinear spectra in the detection frequency domain 
\(\nu_t\) as a function of the variable pump-probe delays \(\tau\), in order to experimentally evaluate the 
dynamics of the band population under high THz pump fields and its timescales. We report the 
experimental results in Figs.~\ref{fig:spec_and_temp_dec}a,~\ref{fig:spec_and_temp_dec}b,~\ref{fig:spec_and_temp_dec}c, showing the associated spectrograms 
for selected pump levels of 60~kV/cm, 240~kV/cm and 430~kV/cm, respectively. We first 
note the variation in the intensity of the plasma resonance in the spectral region labelled A, 
i.e. for \(1.8 \textrm{ THz} < \nu_t < 2.2 \textrm{ THz}\), at \(\tau > 0\)~ps for increasing strength of the THz excitation. The 
lowest THz pump field induces an excitation of the plasma resonance that has an intensity 
maximum shortly after \(\tau  = 0\)~ps, followed by a fast decay down by 40\% around 
\(\tau = 3.3\)~ps. This 
plasma peak intensity reduction with time can be fitted by a decaying exponential, which 
provides a time constant of \((3.7 \pm 0.6)\)~ps.  Conversely, this decay of the plasma peak intensity 
is not visible for higher THz pump fields, which actually induce an almost stable nonlinear 
spectral response of InSb in the spectral range A after \(\tau > 0.5\)~ps. An intermediate THz pump 
level of 100~kV/cm induces a slower decay, with a characteristic time of \((10 \pm 1)\)~ps, while for 
excitation pulses with peak field above 115~kV/cm the traces are can be fitted with constant 
line in our measured range for \(\tau  > 0\)~ps. 

The nonlinear spectrograms also reveal the progressive enhancement of the high-frequency tail of 
the signal for later pump-probe delays and increasing THz pump peak fields. Besides the 
increased spectral intensity near the 5.7 THz LO phonon frequency, the evolution of the 
broadening of the plasma peak can be tracked in a frequency interval where no intrinsic 
resonances of InSb are present. Considering the detection frequency range B between 3.8~THz and 
4.2~THz (see Fig.~\ref{fig:spec_and_temp_dec}a), we can average the spectral intensity and retrieve its evolution as a 
function of the pump-probe delay. Fig.~\ref{fig:spec_and_temp_dec}e reports the time-dependence for THz pump levels 
between 60~kV/cm and 430~kV/cm, showing remarkable differences with respect to the 
behavior of the intensity of the plasma resonance at \(\nu_t  = 2\)~THz shown in Fig.~\ref{fig:spec_and_temp_dec}d. While no 
remarkable time evolution is present for the lowest THz pump intensity, the traces averaged 
around \(\nu_t  = 4\)~THz at higher pump levels show two different features respectively around 
\(\tau = 0\)~ps and for \(\tau > 0.5\)~ps. The earliest feature appears as a sharp peak in correspondence with the 
arrival of the THz pulse on the material, decreasing again within 0.3~ps after most of the pulse 
energy has been delivered. Higher THz pump levels drive the reflected nonlinear signal at 4~THz 
to grow again for pump-probe delays  \(\tau > 0.5\)~ps. 

This temporal analysis of the high-frequency components around \(\nu_t  = 4\)~THz of the nonlinear 
signal reveals the coexistence of two distinct dynamics over different timescales, in agreement 
with reports on angle-resolved photoemission experiments done with a laser pump energy of the 
order of 1~eV~\cite{10.1103/PhysRevB.91.045201}. The faster dynamics between \(\tau = 0\)~ps and \(\tau = 0.3\)~ps are associated to the THz 
pump-driven electron multiplication and acceleration in the \(\Gamma\) valley of the conduction band, 
while the slower dynamics visible for \(\tau > 0.5\)~ps correlates with increasing contributions from 
intervalley scattering and thermalization of hot electrons. Indeed, the associated long-term 
rise of the high-frequency tail is visible for THz pump peak fields above 150~kV/cm, which 
according to our model accelerate the electrons in the \(\Gamma\) valley to energies increasingly larger 
than 0.6~eV, enabling a growing intervalley scattering with the L valley. It has been reported 
that, in turn, such hot electrons in the L valley induce phonon-assisted impact ionization, 
further enhancing electron populations near the minimum of the conduction band in the \(\Gamma\)
valley~\cite{10.1038/s41598-020-67541-1,10.1038/ncomms1598}. In the case of the largest THz peak field of 430~kV/cm, the two regimes are not 
even clearly distinguishable due to a temporal overlap of the fast and slow impact ionization 
regimes, most likely connected to the very high energies reached by the electrons that rapidly 
enable a very efficient intervalley scattering and a faster secondary impact ionization. This 
separate analysis of the evolution in \(\tau\) of the different spectral components around the plasma 
resonance (\(\nu_t  = 2\)~THz) and in the high-frequency tail (for example, \(\nu_t  = 4\)~THz) for 
different THz pump levels allows us to distinguish the fast impact ionization regime in the \(\Gamma\) 
valley and the contributions of a slower impact ionization associated to a complex intervalley 
dynamics, active mainly for THz peak intensities above 150~kV/cm.

In order to quantitatively assess the fast carrier multiplication effect in InSb from these 
experimental data, we investigate more in detail the nonlinear response to directly retrieve the 
carrier multiplication factor from the experimental measurements in different THz pump 
conditions. To do this we focus on positive values of \(\tau\) that are sufficiently large that the overlap of the two THz pulses is minimal, but are also sufficiently small that the relaxation of the carrier density back to equilibrium is negligible.  Inspection of panels a,b and c of Fig.~\ref{fig:spec_and_temp_dec} suggests that an appropriate range of \(\tau\) is between \(\tau_a = 0.4\)~ps and \(\tau_b = 0.7\)~ps.  We then assume that the weaker probe pulse interacts with the sample in a linear optics approximation, with an isotropic reflectivity that is quasi-statically modified by the stronger pump. In this approximation the frequency-dependent reflectivity at normal incidence is
\begin{equation}
    r_j(\nu) = \frac{1 - \sqrt{\epsilon_j(\nu)}}{1+\sqrt{\epsilon_j(\nu)}} 
\end{equation}
where the index \(j\) indicates the state of the material:  \(j = 0\) in equilibrium (with the OR pump pulse blocked) and \(j = 1\) for \(\tau_a < \tau <\tau_b\). Here the permittivity \(\epsilon_j\) can be approximated using a Drude-Lorentz model~\cite{10.1364/OE.27.010854}
\begin{equation}
    \epsilon_j(\nu) = \epsilon_\infty \left( 1+ \frac{\nu_\textrm{LO}^2 - \nu_\textrm{TO}^2}{\nu_\textrm{LO}^2-\nu^2-i2 \pi \nu/\tau_\textrm{ph}} - \frac{\nu_{\textrm{pl},j}^2}{\nu^2 + i 2 \pi \nu / \tau_{\textrm{pl},j}}\right)
\end{equation}
where we assume that 
\(\epsilon_\infty\), \(\nu_\textrm{LO}\), \(\nu_\textrm{TO}\) and \(\tau_\textrm{ph}\) are constants.
In the Drude-Lorentz model the carrier density is proportional to \(\nu_{\textrm{ph},j}^2\) (assuming constant effective mass), and so the carrier multiplication factor can be estimated as
\begin{equation}
    F = \frac{\nu_{\textrm{pl},1}^2}{\nu_{\textrm{pl},0}^2}.\label{eq:Fprobe}
\end{equation}
This implies that we can estimate the carrier multiplication factor \(F\) by measuring the value of the excited-state plasma frequency \(\nu_{\textrm{pl},1}\).

\begin{figure}
    \centering
    \includegraphics[width=14cm]{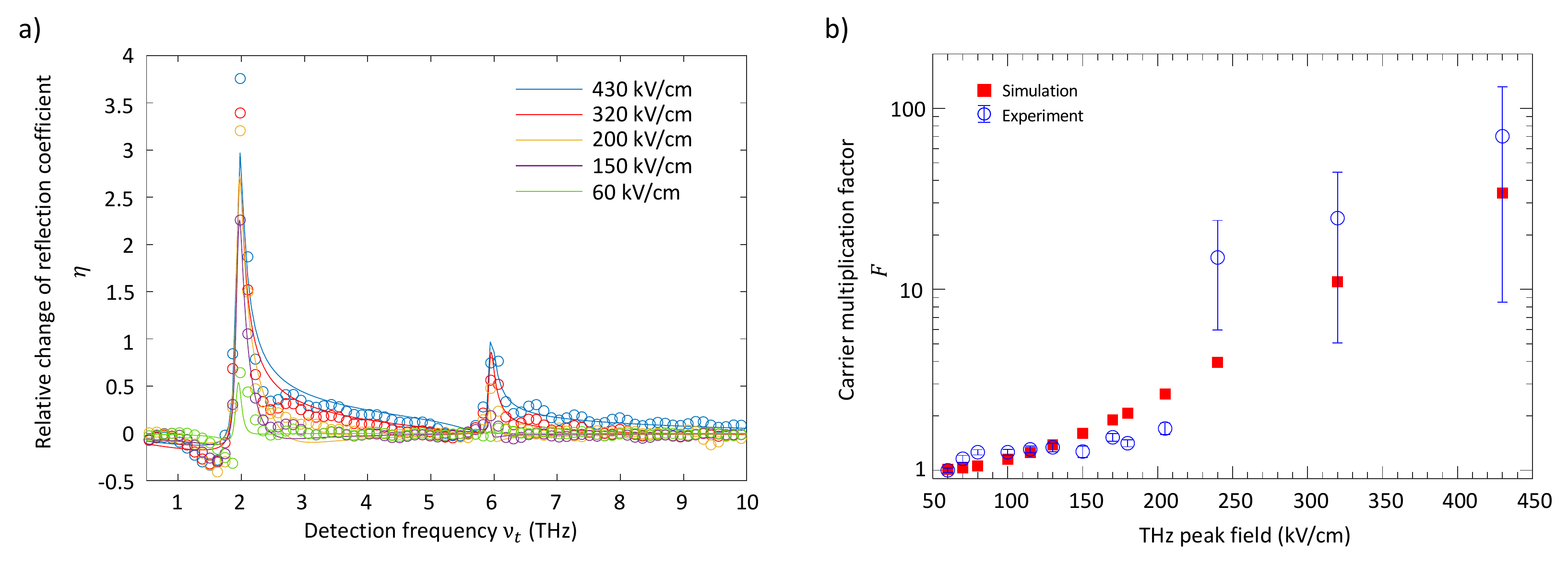}
    \caption{
a) Plot of the relative change \(\eta\) in the magnitude of the reflectivity spectrum, as defined in Eq.~\ref{eq:eta}. The circle markers represent the experimental data points, while the continuous lines are the associated fit results for \(E_\textrm{pump, max}\) ranging from 60 kV/cm to 430 kV/cm. b) Comparison of the carrier multiplication factor \(F\) as function of the peak field of the OR-generated pump pulse, as obtained from the fit of the experimental excited plasma frequency (blue circles) and from the simulations (red squares). 
} \label{fig:Diff_nonlin_probe}
\end{figure}

To extract a value of \(\nu_{\textrm{ph},1}\) from our data, we first define
\begin{equation}
    E_\textrm{probe, exc}^\prime(t) \equiv \frac{1}{\tau_b - \tau_a} \int_{\tau_a}^{\tau_b} E_\textrm{NL probe}^\prime(t,\tau)d\tau
\end{equation}
as the average reflected nonlinear probe field over the interval \(\tau_a < \tau < \tau_b\).  
Let \(\hat{E}^\prime_\textrm{plasma, exc}(\nu)\) be the Fourier transform of \(E_\textrm{probe, exc}^\prime(t)\).  
In our approximation of a 
quasi-equilibrium transient state and linear probe interaction, 
\begin{equation}
    \hat{E}_\textrm{probe, exc}^\prime(\nu) = r_1(\nu) \hat{E}_\textrm{probe}(\nu)
\end{equation}
and
\begin{equation}
    \hat{E}_\textrm{probe}^\prime(\nu) = r_0(\nu) \hat{E}_\textrm{probe}(\nu).
\end{equation}
where \(\hat{E}_\textrm{probe}(\nu)\) and \(\hat{E}_\textrm{probe}^\prime(\nu)\) are the Fourier transforms of the plasma-generated probe pulse electric field incident on and reflected from the sample, respectively.
For fitting to the experimental data we define  the relative change in the magnitude of the probe spectrum as
\begin{equation}
    \eta(\nu) \equiv \frac{|\hat{E}_\textrm{probe, exc}^\prime(\nu)| - |\hat{E}_\textrm{probe}^\prime(\nu)|}{|\hat{E}_\textrm{probe}^\prime(\nu)|} = \frac{|r_1(\nu)| - |r_0(\nu)|}{|r_0(\nu)|}\label{eq:eta}.
\end{equation}
  Fig.~\ref{fig:Diff_nonlin_probe}a shows the behavior of \(\eta(\nu)\) for various values of the peak \(E_\textrm{pump}\) field, as well as fits from the Drude-Lorentz model performed by optimizing the values of \(\nu_{\textrm{ph},j}\) and \(\tau_{\textrm{ph},j}\).  The are marked features at and around 2~THz and 6~THz are reproduced by the model.  As expected, the parameters for the equilibrium free carriers \(\nu_{\textrm{pl},0}\) and  \(\tau_{\textrm{pl},0}\) were highly consistent for all data sets and give average values of \(\nu_{\textrm{pl},0} = (1.97 \pm 0.02)\)~THz and \(\tau_{\textrm{pl},0} = (1.2 \pm 0.3) \)~ps.  The fits shown in Fig.~\ref{fig:Diff_nonlin_probe} have these equilibrium values fixed at these values for all data sets, but with the quasi-equilibrium  parameters still allowed to vary \(\nu_{\textrm{pl},1}\) and  \(\tau_{\textrm{pl},1}\).
The fitted carrier 
relaxation time \(\tau_{\textrm{pl},1}\) lies in the range 0.8~ps and 1.6~ps for THz peak fields below 200~kV/cm, essentially equivalent to the equilibrium value.  At higher fields the fitted scattering time dramatically decreases, reaching a minimum of  \(\tau_{\textrm{pl},1} = 4\)~fs for the strongest THz pump intensity. This is compatible with the timescales of intervalley scattering between the L 
and \(\Gamma\) valleys as well as electron-hole recombination effects~\cite{10.1103/PhysRevB.91.045201}. This suggests the 
coexistence of impact ionization and inter-valley scattering for such high THz pump levels, 
which is compatible with the electron kinetic energy computed with our simulations.

The fitted values of the excited-state plasma frequency \(\nu_{\textrm{pl},1}\) can be directly converted to estimates of the carrier multiplication factor \(F\) using Eq.~\ref{eq:Fprobe}.  The results are shown in  Fig.~\ref{fig:Diff_nonlin_probe}b  as a function of the THz pump peak field.  At low pump fields (below 210~kV) the value of \(F\) is modest, increasing by a factor of less than 2.  For higher fields the quality of the fit is lower, leading to larger uncertainties in \(F\) but nonetheless trending to much larger values, with a value of \(70 \pm 60\) at the maximum field strength of  430~kV/cm. We can compare these 
experimental results with the carrier concentration increase \(F_\textrm{sim}\) predicted by the simulations, 
averaged over the same pump-probe delay interval. As shown in Fig~\ref{fig:Diff_nonlin_probe}b 
(red dots), the simulated values \(F_\textrm{sim}\) are in good agreement with the experimental 
trend for different THz pump levels. We can deduce from this that our simulation framework 
provides a reasonable approximation of the InSb response in presence of significant THz-driven fast 
impact ionization contributions for THz pump peak fields between 60~kV/cm and 430~kV/cm. 

Although the agreement here is limited by fit uncertainties, in principle our prediction of \(F\) might be improved be relaxing some of the assumptions used in the simulation. According to our model, conduction carriers accelerated by fields above 150~kV/cm fields in 
the \(\Gamma\) valley have enough energy to undergo intervalley scattering to the L valley (600~meV), 
while electron energies larger than 1~eV are reached for pump fields of 320~kV/cm and more, 
enabling scattering to the X valley~\cite{10.1038/s41598-020-67541-1,10.1038/ncomms1598}. This would also influence the electron energy 
relaxation time \(\tau_\epsilon\), which was in our model kept constant for simplicity. These dynamic effects 
could be accounted for with additional sets of coupled rate equations and Monte-Carlo modeling, 
at the cost of increased computation complexity and time. Moreover, the formula for the 
experimental assessment of the plasma frequency is based on the effective mass 
approximation, 
which is not very accurate when the electrons populate regions of the conduction band with strong 
non-parabolic features and valleys away from the \(\Gamma\) point.

\section{\label{sec:Conclusison}Conclusion}

We have shown using 2D THz TDS that intense ultrafast THz pump pulses between 60~kV/cm and 430~kV/cm on InSb induce significant carrier multiplication within the first picoseconds. Broadband THz probe pulses are sensitive not only to the resonances associated with the intrinsic plasma and phonon frequencies, but also to changes in the conduction band population. We have also introduced a computationally efficient model for this interaction that reproduces the main features of the experiment, incorporating a macroscopic treatment for impact ionization in addition to previously developed methods for treating ballistic transport. 

More specifically, the simulated 2D nonlinear spectra show good agreement with the experimental results, reporting comparable spectral intensity of the main plasma resonance peak, up to \( 50\)\% of the probe spectrum maximum for the largest THz pump peak field of 430~kV/cm. The model also predicts the enhancement of the high-frequency components of the 2D nonlinear spectra, deriving from the modulation of the plasma frequency induced by the different THz pump level. The analysis of the nonlinear spectra for different pump-probe delays highlights the presence of two timescales in the InSb response evolution, associated with fast impact ionization induced directly by the THz pump pulse and with an additional slow impact ionization effect that becomes dominant for \( \tau > 0.5\)~ps approximately. 

Moreover, the increase of the conduction band population is quantitatively assessed from the fit of the variation of the nonlinear reflection coefficient induced by the different THz pump levels. We retrieve a maximum carrier multiplication factor \(>10\), in good agreement with the fast impact ionization simulation, which further grows when considering later pump-probe intervals with increasing contributions from the slow impact ionization.

Overall, the combination of our 2D THz experiments and the developed hybrid model gives access to the relevant ultrafast timescales and magnitude of the impact ionization process in InSb for a broad range of THz pump conditions. This detailed insight into the ultrafast dynamics at the onset of impact ionization process helps to understand and predict the operation conditions under which InSb-based devices can undergo a transition between the purely ballistic transport and the carrier multiplication regime. Moreover, the presented numerical and experimental methods are a promising platform for the investigation of highly nonlinear effects in the electron dynamics not only in conventional low-bandgap semiconductors, but also in gapless materials possessing exotic band structures like topological semimetals, where impact ionization plays a relevant role already at very low incident THz fields. 

\section*{\label{sec:Ack}Acknowledgments}

We acknowledge Georg Winkler and Quansheng Wu, formerly at the Institute for Theoretical Physics and Station Q at ETH Zurich, for the band structure calculations used in this work.  This research was supported by the NCCR MUST, funded by the Swiss National Science Foundation. S. B. acknowledges that this project has received funding from European Union’s Horizon 2020 under MCSA Grant No. 801459, FP-RESOMUS. E. A. acknowledges support from the Swiss National Science Foundation through Ambizione Grant PZ00P2 179691.

\bibliography{export_2022-8-5}

\begin{thebibliography}{39}%
\makeatletter
\providecommand \@ifxundefined [1]{%
 \@ifx{#1\undefined}
}%
\providecommand \@ifnum [1]{%
 \ifnum #1\expandafter \@firstoftwo
 \else \expandafter \@secondoftwo
 \fi
}%
\providecommand \@ifx [1]{%
 \ifx #1\expandafter \@firstoftwo
 \else \expandafter \@secondoftwo
 \fi
}%
\providecommand \natexlab [1]{#1}%
\providecommand \enquote  [1]{``#1''}%
\providecommand \bibnamefont  [1]{#1}%
\providecommand \bibfnamefont [1]{#1}%
\providecommand \citenamefont [1]{#1}%
\providecommand \href@noop [0]{\@secondoftwo}%
\providecommand \href [0]{\begingroup \@sanitize@url \@href}%
\providecommand \@href[1]{\@@startlink{#1}\@@href}%
\providecommand \@@href[1]{\endgroup#1\@@endlink}%
\providecommand \@sanitize@url [0]{\catcode `\\12\catcode `\$12\catcode
  `\&12\catcode `\#12\catcode `\^12\catcode `\_12\catcode `\%12\relax}%
\providecommand \@@startlink[1]{}%
\providecommand \@@endlink[0]{}%
\providecommand \url  [0]{\begingroup\@sanitize@url \@url }%
\providecommand \@url [1]{\endgroup\@href {#1}{\urlprefix }}%
\providecommand \urlprefix  [0]{URL }%
\providecommand \Eprint [0]{\href }%
\providecommand \doibase [0]{https://doi.org/}%
\providecommand \selectlanguage [0]{\@gobble}%
\providecommand \bibinfo  [0]{\@secondoftwo}%
\providecommand \bibfield  [0]{\@secondoftwo}%
\providecommand \translation [1]{[#1]}%
\providecommand \BibitemOpen [0]{}%
\providecommand \bibitemStop [0]{}%
\providecommand \bibitemNoStop [0]{.\EOS\space}%
\providecommand \EOS [0]{\spacefactor3000\relax}%
\providecommand \BibitemShut  [1]{\csname bibitem#1\endcsname}%
\let\auto@bib@innerbib\@empty
\bibitem [{\citenamefont {Tonouchi}(2007)}]{10.1038/nphoton.2007.3}%
  \BibitemOpen
  \bibfield  {author} {\bibinfo {author} {\bibfnamefont {M.}~\bibnamefont
  {Tonouchi}},\ }\bibfield  {title} {\bibinfo {title} {{Cutting-edge terahertz
  technology}},\ }\href {https://doi.org/10.1038/nphoton.2007.3} {\bibfield
  {journal} {\bibinfo  {journal} {Nature Photonics}\ }\textbf {\bibinfo
  {volume} {1}},\ \bibinfo {pages} {97} (\bibinfo {year} {2007})}\BibitemShut
  {NoStop}%
\bibitem [{\citenamefont {Williams}(2006)}]{10.1088/0034-4885/69/2/r01}%
  \BibitemOpen
  \bibfield  {author} {\bibinfo {author} {\bibfnamefont {G.~P.}\ \bibnamefont
  {Williams}},\ }\bibfield  {title} {\bibinfo {title} {{Filling the THz
  gap—high power sources and applications}},\ }\href
  {https://doi.org/10.1088/0034-4885/69/2/r01} {\bibfield  {journal} {\bibinfo
  {journal} {Reports on Progress in Physics}\ }\textbf {\bibinfo {volume}
  {69}},\ \bibinfo {pages} {301} (\bibinfo {year} {2006})}\BibitemShut
  {NoStop}%
\bibitem [{\citenamefont {Kuehn}\ \emph
  {et~al.}(2010{\natexlab{a}})\citenamefont {Kuehn}, \citenamefont {Gaal},
  \citenamefont {Reimann}, \citenamefont {Woerner}, \citenamefont {Elsaesser},\
  and\ \citenamefont {Hey}}]{10.1103/PhysRevLett.104.146602}%
  \BibitemOpen
  \bibfield  {author} {\bibinfo {author} {\bibfnamefont {W.}~\bibnamefont
  {Kuehn}}, \bibinfo {author} {\bibfnamefont {P.}~\bibnamefont {Gaal}},
  \bibinfo {author} {\bibfnamefont {K.}~\bibnamefont {Reimann}}, \bibinfo
  {author} {\bibfnamefont {M.}~\bibnamefont {Woerner}}, \bibinfo {author}
  {\bibfnamefont {T.}~\bibnamefont {Elsaesser}},\ and\ \bibinfo {author}
  {\bibfnamefont {R.}~\bibnamefont {Hey}},\ }\bibfield  {title} {\bibinfo
  {title} {{Coherent Ballistic Motion of Electrons in a Periodic Potential}},\
  }\href {https://doi.org/10.1103/physrevlett.104.146602} {\bibfield  {journal}
  {\bibinfo  {journal} {Physical Review Letters}\ }\textbf {\bibinfo {volume}
  {104}},\ \bibinfo {pages} {146602} (\bibinfo {year}
  {2010}{\natexlab{a}})}\BibitemShut {NoStop}%
\bibitem [{\citenamefont {Blanchard}\ \emph {et~al.}(2011)\citenamefont
  {Blanchard}, \citenamefont {Golde}, \citenamefont {Su}, \citenamefont
  {Razzari}, \citenamefont {Sharma}, \citenamefont {Morandotti}, \citenamefont
  {Ozaki}, \citenamefont {Reid}, \citenamefont {Kira}, \citenamefont {Koch},\
  and\ \citenamefont {Hegmann}}]{10.1103/PhysRevLett.107.107401}%
  \BibitemOpen
  \bibfield  {author} {\bibinfo {author} {\bibfnamefont {F.}~\bibnamefont
  {Blanchard}}, \bibinfo {author} {\bibfnamefont {D.}~\bibnamefont {Golde}},
  \bibinfo {author} {\bibfnamefont {F.~H.}\ \bibnamefont {Su}}, \bibinfo
  {author} {\bibfnamefont {L.}~\bibnamefont {Razzari}}, \bibinfo {author}
  {\bibfnamefont {G.}~\bibnamefont {Sharma}}, \bibinfo {author} {\bibfnamefont
  {R.}~\bibnamefont {Morandotti}}, \bibinfo {author} {\bibfnamefont
  {T.}~\bibnamefont {Ozaki}}, \bibinfo {author} {\bibfnamefont
  {M.}~\bibnamefont {Reid}}, \bibinfo {author} {\bibfnamefont {M.}~\bibnamefont
  {Kira}}, \bibinfo {author} {\bibfnamefont {S.~W.}\ \bibnamefont {Koch}},\
  and\ \bibinfo {author} {\bibfnamefont {F.~A.}\ \bibnamefont {Hegmann}},\
  }\bibfield  {title} {\bibinfo {title} {{Effective Mass Anisotropy of Hot
  Electrons in Nonparabolic Conduction Bands of n-Doped InGaAs Films Using
  Ultrafast Terahertz Pump-Probe Techniques}},\ }\href
  {https://doi.org/10.1103/physrevlett.107.107401} {\bibfield  {journal}
  {\bibinfo  {journal} {Physical Review Letters}\ }\textbf {\bibinfo {volume}
  {107}},\ \bibinfo {pages} {107401} (\bibinfo {year} {2011})}\BibitemShut
  {NoStop}%
\bibitem [{\citenamefont {Kuehn}\ \emph
  {et~al.}(2010{\natexlab{b}})\citenamefont {Kuehn}, \citenamefont {Gaal},
  \citenamefont {Reimann}, \citenamefont {Woerner}, \citenamefont {Elsaesser},\
  and\ \citenamefont {Hey}}]{10.1103/PhysRevB.82.075204}%
  \BibitemOpen
  \bibfield  {author} {\bibinfo {author} {\bibfnamefont {W.}~\bibnamefont
  {Kuehn}}, \bibinfo {author} {\bibfnamefont {P.}~\bibnamefont {Gaal}},
  \bibinfo {author} {\bibfnamefont {K.}~\bibnamefont {Reimann}}, \bibinfo
  {author} {\bibfnamefont {M.}~\bibnamefont {Woerner}}, \bibinfo {author}
  {\bibfnamefont {T.}~\bibnamefont {Elsaesser}},\ and\ \bibinfo {author}
  {\bibfnamefont {R.}~\bibnamefont {Hey}},\ }\bibfield  {title} {\bibinfo
  {title} {{Terahertz-induced interband tunneling of electrons in GaAs}},\
  }\href {https://doi.org/10.1103/physrevb.82.075204} {\bibfield  {journal}
  {\bibinfo  {journal} {Physical Review B}\ }\textbf {\bibinfo {volume} {82}},\
  \bibinfo {pages} {075204} (\bibinfo {year} {2010}{\natexlab{b}})}\BibitemShut
  {NoStop}%
\bibitem [{\citenamefont {Hebling}\ \emph {et~al.}(2010)\citenamefont
  {Hebling}, \citenamefont {Hoffmann}, \citenamefont {Hwang}, \citenamefont
  {Yeh},\ and\ \citenamefont {Nelson}}]{10.1103/PhysRevB.81.035201}%
  \BibitemOpen
  \bibfield  {author} {\bibinfo {author} {\bibfnamefont {J.}~\bibnamefont
  {Hebling}}, \bibinfo {author} {\bibfnamefont {M.~C.}\ \bibnamefont
  {Hoffmann}}, \bibinfo {author} {\bibfnamefont {H.~Y.}\ \bibnamefont {Hwang}},
  \bibinfo {author} {\bibfnamefont {K.-L.}\ \bibnamefont {Yeh}},\ and\ \bibinfo
  {author} {\bibfnamefont {K.~A.}\ \bibnamefont {Nelson}},\ }\bibfield  {title}
  {\bibinfo {title} {{Observation of nonequilibrium carrier distribution in Ge,
  Si, and GaAs by terahertz pump–terahertz probe measurements}},\ }\href
  {https://doi.org/10.1103/physrevb.81.035201} {\bibfield  {journal} {\bibinfo
  {journal} {Physical Review B}\ }\textbf {\bibinfo {volume} {81}},\ \bibinfo
  {pages} {035201} (\bibinfo {year} {2010})}\BibitemShut {NoStop}%
\bibitem [{\citenamefont {Sharma}\ \emph {et~al.}(2010)\citenamefont {Sharma},
  \citenamefont {Razzari}, \citenamefont {Su}, \citenamefont {Blanchard},
  \citenamefont {Ayesheshim}, \citenamefont {Cocker}, \citenamefont {Titova},
  \citenamefont {Bandulet}, \citenamefont {Ozaki}, \citenamefont {Kieffer},
  \citenamefont {Morandotti}, \citenamefont {Reid},\ and\ \citenamefont
  {Hegmann}}]{10.1109/JPHOT.2010.2050873}%
  \BibitemOpen
  \bibfield  {author} {\bibinfo {author} {\bibfnamefont {G.}~\bibnamefont
  {Sharma}}, \bibinfo {author} {\bibfnamefont {L.}~\bibnamefont {Razzari}},
  \bibinfo {author} {\bibfnamefont {F.~H.}\ \bibnamefont {Su}}, \bibinfo
  {author} {\bibfnamefont {F.}~\bibnamefont {Blanchard}}, \bibinfo {author}
  {\bibfnamefont {A.}~\bibnamefont {Ayesheshim}}, \bibinfo {author}
  {\bibfnamefont {T.~L.}\ \bibnamefont {Cocker}}, \bibinfo {author}
  {\bibfnamefont {L.~V.}\ \bibnamefont {Titova}}, \bibinfo {author}
  {\bibfnamefont {H.~C.}\ \bibnamefont {Bandulet}}, \bibinfo {author}
  {\bibfnamefont {T.}~\bibnamefont {Ozaki}}, \bibinfo {author} {\bibfnamefont
  {J.-C.}\ \bibnamefont {Kieffer}}, \bibinfo {author} {\bibfnamefont
  {R.}~\bibnamefont {Morandotti}}, \bibinfo {author} {\bibfnamefont
  {M.}~\bibnamefont {Reid}},\ and\ \bibinfo {author} {\bibfnamefont {F.~A.}\
  \bibnamefont {Hegmann}},\ }\bibfield  {title} {\bibinfo {title}
  {{Time-Resolved Terahertz Spectroscopy of Free Carrier Nonlinear Dynamics in
  Semiconductors}},\ }\href {https://doi.org/10.1109/jphot.2010.2050873}
  {\bibfield  {journal} {\bibinfo  {journal} {IEEE Photonics Journal}\ }\textbf
  {\bibinfo {volume} {2}},\ \bibinfo {pages} {578} (\bibinfo {year}
  {2010})}\BibitemShut {NoStop}%
\bibitem [{\citenamefont {Wen}\ \emph {et~al.}(2008)\citenamefont {Wen},
  \citenamefont {Wiczer},\ and\ \citenamefont
  {Lindenberg}}]{10.1103/PhysRevB.78.125203}%
  \BibitemOpen
  \bibfield  {author} {\bibinfo {author} {\bibfnamefont {H.}~\bibnamefont
  {Wen}}, \bibinfo {author} {\bibfnamefont {M.}~\bibnamefont {Wiczer}},\ and\
  \bibinfo {author} {\bibfnamefont {A.~M.}\ \bibnamefont {Lindenberg}},\
  }\bibfield  {title} {\bibinfo {title} {{Ultrafast electron cascades in
  semiconductors driven by intense femtosecond terahertz pulses}},\ }\href
  {https://doi.org/10.1103/physrevb.78.125203} {\bibfield  {journal} {\bibinfo
  {journal} {Physical Review B}\ }\textbf {\bibinfo {volume} {78}},\ \bibinfo
  {pages} {125203} (\bibinfo {year} {2008})}\BibitemShut {NoStop}%
\bibitem [{\citenamefont {Hoffmann}\ \emph {et~al.}(2009)\citenamefont
  {Hoffmann}, \citenamefont {Hebling}, \citenamefont {Hwang}, \citenamefont
  {Yeh},\ and\ \citenamefont {Nelson}}]{10.1103/PhysRevB.79.161201}%
  \BibitemOpen
  \bibfield  {author} {\bibinfo {author} {\bibfnamefont {M.~C.}\ \bibnamefont
  {Hoffmann}}, \bibinfo {author} {\bibfnamefont {J.}~\bibnamefont {Hebling}},
  \bibinfo {author} {\bibfnamefont {H.~Y.}\ \bibnamefont {Hwang}}, \bibinfo
  {author} {\bibfnamefont {K.-L.}\ \bibnamefont {Yeh}},\ and\ \bibinfo {author}
  {\bibfnamefont {K.~A.}\ \bibnamefont {Nelson}},\ }\bibfield  {title}
  {\bibinfo {title} {{Impact ionization in InSb probed by terahertz
  pump—terahertz probe spectroscopy}},\ }\href
  {https://doi.org/10.1103/physrevb.79.161201} {\bibfield  {journal} {\bibinfo
  {journal} {Physical Review B}\ }\textbf {\bibinfo {volume} {79}},\ \bibinfo
  {pages} {161201} (\bibinfo {year} {2009})}\BibitemShut {NoStop}%
\bibitem [{\citenamefont {Woerner}\ \emph {et~al.}(2013)\citenamefont
  {Woerner}, \citenamefont {Kuehn}, \citenamefont {Bowlan}, \citenamefont
  {Reimann},\ and\ \citenamefont {Elsaesser}}]{10.1088/1367-2630/15/2/025039}%
  \BibitemOpen
  \bibfield  {author} {\bibinfo {author} {\bibfnamefont {M.}~\bibnamefont
  {Woerner}}, \bibinfo {author} {\bibfnamefont {W.}~\bibnamefont {Kuehn}},
  \bibinfo {author} {\bibfnamefont {P.}~\bibnamefont {Bowlan}}, \bibinfo
  {author} {\bibfnamefont {K.}~\bibnamefont {Reimann}},\ and\ \bibinfo {author}
  {\bibfnamefont {T.}~\bibnamefont {Elsaesser}},\ }\bibfield  {title} {\bibinfo
  {title} {{Ultrafast two-dimensional terahertz spectroscopy of elementary
  excitations in solids}},\ }\href
  {https://doi.org/10.1088/1367-2630/15/2/025039} {\bibfield  {journal}
  {\bibinfo  {journal} {New Journal of Physics}\ }\textbf {\bibinfo {volume}
  {15}},\ \bibinfo {pages} {025039} (\bibinfo {year} {2013})}\BibitemShut
  {NoStop}%
\bibitem [{\citenamefont {Kuehn}\ \emph {et~al.}(2011)\citenamefont {Kuehn},
  \citenamefont {Reimann}, \citenamefont {Woerner}, \citenamefont {Elsaesser},
  \citenamefont {Hey},\ and\ \citenamefont
  {Schade}}]{10.1103/PhysRevLett.107.067401}%
  \BibitemOpen
  \bibfield  {author} {\bibinfo {author} {\bibfnamefont {W.}~\bibnamefont
  {Kuehn}}, \bibinfo {author} {\bibfnamefont {K.}~\bibnamefont {Reimann}},
  \bibinfo {author} {\bibfnamefont {M.}~\bibnamefont {Woerner}}, \bibinfo
  {author} {\bibfnamefont {T.}~\bibnamefont {Elsaesser}}, \bibinfo {author}
  {\bibfnamefont {R.}~\bibnamefont {Hey}},\ and\ \bibinfo {author}
  {\bibfnamefont {U.}~\bibnamefont {Schade}},\ }\bibfield  {title} {\bibinfo
  {title} {{Strong Correlation of Electronic and Lattice Excitations in
  GaAs/AlGaAs Semiconductor Quantum Wells Revealed by Two-Dimensional Terahertz
  Spectroscopy}},\ }\href {https://doi.org/10.1103/physrevlett.107.067401}
  {\bibfield  {journal} {\bibinfo  {journal} {Physical Review Letters}\
  }\textbf {\bibinfo {volume} {107}},\ \bibinfo {pages} {067401} (\bibinfo
  {year} {2011})}\BibitemShut {NoStop}%
\bibitem [{\citenamefont {Somma}\ \emph {et~al.}(2016)\citenamefont {Somma},
  \citenamefont {Folpini}, \citenamefont {Reimann}, \citenamefont {Woerner},\
  and\ \citenamefont {Elsaesser}}]{10.1103/PhysRevLett.116.177401}%
  \BibitemOpen
  \bibfield  {author} {\bibinfo {author} {\bibfnamefont {C.}~\bibnamefont
  {Somma}}, \bibinfo {author} {\bibfnamefont {G.}~\bibnamefont {Folpini}},
  \bibinfo {author} {\bibfnamefont {K.}~\bibnamefont {Reimann}}, \bibinfo
  {author} {\bibfnamefont {M.}~\bibnamefont {Woerner}},\ and\ \bibinfo {author}
  {\bibfnamefont {T.}~\bibnamefont {Elsaesser}},\ }\bibfield  {title} {\bibinfo
  {title} {{Two-Phonon Quantum Coherences in Indium Antimonide Studied by
  Nonlinear Two-Dimensional Terahertz Spectroscopy}},\ }\href
  {https://doi.org/10.1103/physrevlett.116.177401} {\bibfield  {journal}
  {\bibinfo  {journal} {Physical Review Letters}\ }\textbf {\bibinfo {volume}
  {116}},\ \bibinfo {pages} {177401} (\bibinfo {year} {2016})}\BibitemShut
  {NoStop}%
\bibitem [{\citenamefont {Pal}\ \emph {et~al.}(2021)\citenamefont {Pal},
  \citenamefont {Strkalj}, \citenamefont {Yang}, \citenamefont {Weber},
  \citenamefont {Trassin}, \citenamefont {Woerner},\ and\ \citenamefont
  {Fiebig}}]{10.1103/PhysRevX.11.021023}%
  \BibitemOpen
  \bibfield  {author} {\bibinfo {author} {\bibfnamefont {S.}~\bibnamefont
  {Pal}}, \bibinfo {author} {\bibfnamefont {N.}~\bibnamefont {Strkalj}},
  \bibinfo {author} {\bibfnamefont {C.-J.}\ \bibnamefont {Yang}}, \bibinfo
  {author} {\bibfnamefont {M.~C.}\ \bibnamefont {Weber}}, \bibinfo {author}
  {\bibfnamefont {M.}~\bibnamefont {Trassin}}, \bibinfo {author} {\bibfnamefont
  {M.}~\bibnamefont {Woerner}},\ and\ \bibinfo {author} {\bibfnamefont
  {M.}~\bibnamefont {Fiebig}},\ }\bibfield  {title} {\bibinfo {title} {{Origin
  of Terahertz Soft-Mode Nonlinearities in Ferroelectric Perovskites}},\ }\href
  {https://doi.org/10.1103/physrevx.11.021023} {\bibfield  {journal} {\bibinfo
  {journal} {Physical Review X}\ }\textbf {\bibinfo {volume} {11}},\ \bibinfo
  {pages} {021023} (\bibinfo {year} {2021})}\BibitemShut {NoStop}%
\bibitem [{\citenamefont {Ting}\ \emph {et~al.}(2019)\citenamefont {Ting},
  \citenamefont {Soibel}, \citenamefont {Khoshakhlagh}, \citenamefont {Keo},
  \citenamefont {Rafol}, \citenamefont {Fisher}, \citenamefont {Pepper},
  \citenamefont {Luong}, \citenamefont {Hill},\ and\ \citenamefont
  {Gunapala}}]{10.1016/j.infrared.2018.12.034}%
  \BibitemOpen
  \bibfield  {author} {\bibinfo {author} {\bibfnamefont {D.~Z.}\ \bibnamefont
  {Ting}}, \bibinfo {author} {\bibfnamefont {A.}~\bibnamefont {Soibel}},
  \bibinfo {author} {\bibfnamefont {A.}~\bibnamefont {Khoshakhlagh}}, \bibinfo
  {author} {\bibfnamefont {S.~A.}\ \bibnamefont {Keo}}, \bibinfo {author}
  {\bibfnamefont {S.~B.}\ \bibnamefont {Rafol}}, \bibinfo {author}
  {\bibfnamefont {A.~M.}\ \bibnamefont {Fisher}}, \bibinfo {author}
  {\bibfnamefont {B.~J.}\ \bibnamefont {Pepper}}, \bibinfo {author}
  {\bibfnamefont {E.~M.}\ \bibnamefont {Luong}}, \bibinfo {author}
  {\bibfnamefont {C.~J.}\ \bibnamefont {Hill}},\ and\ \bibinfo {author}
  {\bibfnamefont {S.~D.}\ \bibnamefont {Gunapala}},\ }\bibfield  {title}
  {\bibinfo {title} {{Advances in III-V semiconductor infrared absorbers and
  detectors}},\ }\href {https://doi.org/10.1016/j.infrared.2018.12.034}
  {\bibfield  {journal} {\bibinfo  {journal} {Infrared Physics \& Technology}\
  }\textbf {\bibinfo {volume} {97}},\ \bibinfo {pages} {210} (\bibinfo {year}
  {2019})}\BibitemShut {NoStop}%
\bibitem [{\citenamefont {Kanno}\ \emph {et~al.}(2007)\citenamefont {Kanno},
  \citenamefont {Hishiki}, \citenamefont {Kogetsu}, \citenamefont {Nakamura},\
  and\ \citenamefont {Katagiri}}]{10.1063/1.2737768}%
  \BibitemOpen
  \bibfield  {author} {\bibinfo {author} {\bibfnamefont {I.}~\bibnamefont
  {Kanno}}, \bibinfo {author} {\bibfnamefont {S.}~\bibnamefont {Hishiki}},
  \bibinfo {author} {\bibfnamefont {Y.}~\bibnamefont {Kogetsu}}, \bibinfo
  {author} {\bibfnamefont {T.}~\bibnamefont {Nakamura}},\ and\ \bibinfo
  {author} {\bibfnamefont {M.}~\bibnamefont {Katagiri}},\ }\bibfield  {title}
  {\bibinfo {title} {{Fast response of InSb Schottky detector}},\ }\href
  {https://doi.org/10.1063/1.2737768} {\bibfield  {journal} {\bibinfo
  {journal} {Review of Scientific Instruments}\ }\textbf {\bibinfo {volume}
  {78}},\ \bibinfo {pages} {056103} (\bibinfo {year} {2007})}\BibitemShut
  {NoStop}%
\bibitem [{\citenamefont {Ashley}\ \emph {et~al.}(1995)\citenamefont {Ashley},
  \citenamefont {Dean}, \citenamefont {Elliott}, \citenamefont {Pryce},
  \citenamefont {Johnson},\ and\ \citenamefont {Willis}}]{10.1063/1.114063}%
  \BibitemOpen
  \bibfield  {author} {\bibinfo {author} {\bibfnamefont {T.}~\bibnamefont
  {Ashley}}, \bibinfo {author} {\bibfnamefont {A.~B.}\ \bibnamefont {Dean}},
  \bibinfo {author} {\bibfnamefont {C.~T.}\ \bibnamefont {Elliott}}, \bibinfo
  {author} {\bibfnamefont {G.~J.}\ \bibnamefont {Pryce}}, \bibinfo {author}
  {\bibfnamefont {A.~D.}\ \bibnamefont {Johnson}},\ and\ \bibinfo {author}
  {\bibfnamefont {H.}~\bibnamefont {Willis}},\ }\bibfield  {title} {\bibinfo
  {title} {{Uncooled high‐speed InSb field‐effect transistors}},\ }\href
  {https://doi.org/10.1063/1.114063} {\bibfield  {journal} {\bibinfo  {journal}
  {Applied Physics Letters}\ }\textbf {\bibinfo {volume} {66}},\ \bibinfo
  {pages} {481} (\bibinfo {year} {1995})}\BibitemShut {NoStop}%
\bibitem [{\citenamefont {Ashley}\ \emph {et~al.}(2007)\citenamefont {Ashley},
  \citenamefont {Buckle}, \citenamefont {Datta}, \citenamefont {Emeny},
  \citenamefont {Hayes}, \citenamefont {Hilton}, \citenamefont {Jefferies},
  \citenamefont {Martin}, \citenamefont {Phillips}, \citenamefont {Wallis},
  \citenamefont {Wilding},\ and\ \citenamefont {Chau}}]{10.1049/el:20071335}%
  \BibitemOpen
  \bibfield  {author} {\bibinfo {author} {\bibfnamefont {T.}~\bibnamefont
  {Ashley}}, \bibinfo {author} {\bibfnamefont {L.}~\bibnamefont {Buckle}},
  \bibinfo {author} {\bibfnamefont {S.}~\bibnamefont {Datta}}, \bibinfo
  {author} {\bibfnamefont {M.~T.}\ \bibnamefont {Emeny}}, \bibinfo {author}
  {\bibfnamefont {D.~G.}\ \bibnamefont {Hayes}}, \bibinfo {author}
  {\bibfnamefont {K.~P.}\ \bibnamefont {Hilton}}, \bibinfo {author}
  {\bibfnamefont {R.}~\bibnamefont {Jefferies}}, \bibinfo {author}
  {\bibfnamefont {T.}~\bibnamefont {Martin}}, \bibinfo {author} {\bibfnamefont
  {T.~J.}\ \bibnamefont {Phillips}}, \bibinfo {author} {\bibfnamefont {D.~J.}\
  \bibnamefont {Wallis}}, \bibinfo {author} {\bibfnamefont {P.~J.}\
  \bibnamefont {Wilding}},\ and\ \bibinfo {author} {\bibfnamefont
  {R.}~\bibnamefont {Chau}},\ }\bibfield  {title} {\bibinfo {title}
  {{Heterogeneous InSb quantum well transistors on silicon for ultra-high
  speed, low power logic applications}},\ }\href
  {https://doi.org/10.1049/el:20071335} {\bibfield  {journal} {\bibinfo
  {journal} {Electronics Letters}\ }\textbf {\bibinfo {volume} {43}},\ \bibinfo
  {pages} {777} (\bibinfo {year} {2007})}\BibitemShut {NoStop}%
\bibitem [{\citenamefont {Nilsson}\ \emph {et~al.}(2010)\citenamefont
  {Nilsson}, \citenamefont {Caroff}, \citenamefont {Thelander}, \citenamefont
  {Lind}, \citenamefont {Karlström},\ and\ \citenamefont
  {Wernersson}}]{10.1063/1.3402760}%
  \BibitemOpen
  \bibfield  {author} {\bibinfo {author} {\bibfnamefont {H.~A.}\ \bibnamefont
  {Nilsson}}, \bibinfo {author} {\bibfnamefont {P.}~\bibnamefont {Caroff}},
  \bibinfo {author} {\bibfnamefont {C.}~\bibnamefont {Thelander}}, \bibinfo
  {author} {\bibfnamefont {E.}~\bibnamefont {Lind}}, \bibinfo {author}
  {\bibfnamefont {O.}~\bibnamefont {Karlström}},\ and\ \bibinfo {author}
  {\bibfnamefont {L.-E.}\ \bibnamefont {Wernersson}},\ }\bibfield  {title}
  {\bibinfo {title} {{Temperature dependent properties of InSb and InAs
  nanowire field-effect transistors}},\ }\href
  {https://doi.org/10.1063/1.3402760} {\bibfield  {journal} {\bibinfo
  {journal} {Applied Physics Letters}\ }\textbf {\bibinfo {volume} {96}},\
  \bibinfo {pages} {153505} (\bibinfo {year} {2010})}\BibitemShut {NoStop}%
\bibitem [{\citenamefont {Mata}\ \emph {et~al.}(2016)\citenamefont {Mata},
  \citenamefont {Leturcq}, \citenamefont {Plissard}, \citenamefont {Rolland},
  \citenamefont {Magén}, \citenamefont {Arbiol},\ and\ \citenamefont
  {Caroff}}]{10.1021/acs.nanolett.5b05125}%
  \BibitemOpen
  \bibfield  {author} {\bibinfo {author} {\bibfnamefont {M.~d.~l.}\
  \bibnamefont {Mata}}, \bibinfo {author} {\bibfnamefont {R.}~\bibnamefont
  {Leturcq}}, \bibinfo {author} {\bibfnamefont {S.~R.}\ \bibnamefont
  {Plissard}}, \bibinfo {author} {\bibfnamefont {C.}~\bibnamefont {Rolland}},
  \bibinfo {author} {\bibfnamefont {C.}~\bibnamefont {Magén}}, \bibinfo
  {author} {\bibfnamefont {J.}~\bibnamefont {Arbiol}},\ and\ \bibinfo {author}
  {\bibfnamefont {P.}~\bibnamefont {Caroff}},\ }\bibfield  {title} {\bibinfo
  {title} {{Twin-Induced InSb Nanosails: A Convenient High Mobility Quantum
  System}},\ }\href {https://doi.org/10.1021/acs.nanolett.5b05125} {\bibfield
  {journal} {\bibinfo  {journal} {Nano Letters}\ }\textbf {\bibinfo {volume}
  {16}},\ \bibinfo {pages} {825} (\bibinfo {year} {2016})}\BibitemShut
  {NoStop}%
\bibitem [{\citenamefont {Litvinenko}\ \emph {et~al.}(2008)\citenamefont
  {Litvinenko}, \citenamefont {Nikzad}, \citenamefont {Pidgeon}, \citenamefont
  {Allam}, \citenamefont {Cohen}, \citenamefont {Ashley}, \citenamefont
  {Emeny}, \citenamefont {Zawadzki},\ and\ \citenamefont
  {Murdin}}]{10.1103/PhysRevB.77.033204}%
  \BibitemOpen
  \bibfield  {author} {\bibinfo {author} {\bibfnamefont {K.~L.}\ \bibnamefont
  {Litvinenko}}, \bibinfo {author} {\bibfnamefont {L.}~\bibnamefont {Nikzad}},
  \bibinfo {author} {\bibfnamefont {C.~R.}\ \bibnamefont {Pidgeon}}, \bibinfo
  {author} {\bibfnamefont {J.}~\bibnamefont {Allam}}, \bibinfo {author}
  {\bibfnamefont {L.~F.}\ \bibnamefont {Cohen}}, \bibinfo {author}
  {\bibfnamefont {T.}~\bibnamefont {Ashley}}, \bibinfo {author} {\bibfnamefont
  {M.}~\bibnamefont {Emeny}}, \bibinfo {author} {\bibfnamefont
  {W.}~\bibnamefont {Zawadzki}},\ and\ \bibinfo {author} {\bibfnamefont
  {B.~N.}\ \bibnamefont {Murdin}},\ }\bibfield  {title} {\bibinfo {title}
  {{Temperature dependence of the electron Landé g factor in InSb and GaAs}},\
  }\href {https://doi.org/10.1103/physrevb.77.033204} {\bibfield  {journal}
  {\bibinfo  {journal} {Physical Review B}\ }\textbf {\bibinfo {volume} {77}},\
  \bibinfo {pages} {033204} (\bibinfo {year} {2008})}\BibitemShut {NoStop}%
\bibitem [{\citenamefont {Littler}\ and\ \citenamefont
  {Seiler}(1985)}]{10.1063/1.95789}%
  \BibitemOpen
  \bibfield  {author} {\bibinfo {author} {\bibfnamefont {C.~L.}\ \bibnamefont
  {Littler}}\ and\ \bibinfo {author} {\bibfnamefont {D.~G.}\ \bibnamefont
  {Seiler}},\ }\bibfield  {title} {\bibinfo {title} {{Temperature dependence of
  the energy gap of InSb using nonlinear optical techniques}},\ }\href
  {https://doi.org/10.1063/1.95789} {\bibfield  {journal} {\bibinfo  {journal}
  {Applied Physics Letters}\ }\textbf {\bibinfo {volume} {46}},\ \bibinfo
  {pages} {986} (\bibinfo {year} {1985})}\BibitemShut {NoStop}%
\bibitem [{\citenamefont {Žutić}\ \emph {et~al.}(2004)\citenamefont
  {Žutić}, \citenamefont {Fabian},\ and\ \citenamefont
  {Sarma}}]{10.1103/RevModPhys.76.323}%
  \BibitemOpen
  \bibfield  {author} {\bibinfo {author} {\bibfnamefont {I.}~\bibnamefont
  {Žutić}}, \bibinfo {author} {\bibfnamefont {J.}~\bibnamefont {Fabian}},\
  and\ \bibinfo {author} {\bibfnamefont {S.~D.}\ \bibnamefont {Sarma}},\
  }\bibfield  {title} {\bibinfo {title} {{Spintronics: Fundamentals and
  applications}},\ }\href {https://doi.org/10.1103/revmodphys.76.323}
  {\bibfield  {journal} {\bibinfo  {journal} {Reviews of Modern Physics}\
  }\textbf {\bibinfo {volume} {76}},\ \bibinfo {pages} {323} (\bibinfo {year}
  {2004})}\BibitemShut {NoStop}%
\bibitem [{\citenamefont {Chochol}\ \emph {et~al.}(2016)\citenamefont
  {Chochol}, \citenamefont {Postava}, \citenamefont {Čada}, \citenamefont
  {Vanwolleghem}, \citenamefont {Halagacka}, \citenamefont {Lampin},\ and\
  \citenamefont {Pištora}}]{10.1063/1.4968178}%
  \BibitemOpen
  \bibfield  {author} {\bibinfo {author} {\bibfnamefont {J.}~\bibnamefont
  {Chochol}}, \bibinfo {author} {\bibfnamefont {K.}~\bibnamefont {Postava}},
  \bibinfo {author} {\bibfnamefont {M.}~\bibnamefont {Čada}}, \bibinfo
  {author} {\bibfnamefont {M.}~\bibnamefont {Vanwolleghem}}, \bibinfo {author}
  {\bibfnamefont {L.}~\bibnamefont {Halagacka}}, \bibinfo {author}
  {\bibfnamefont {J.-F.}\ \bibnamefont {Lampin}},\ and\ \bibinfo {author}
  {\bibfnamefont {J.}~\bibnamefont {Pištora}},\ }\bibfield  {title} {\bibinfo
  {title} {{Magneto-optical properties of InSb for terahertz applications}},\
  }\href {https://doi.org/10.1063/1.4968178} {\bibfield  {journal} {\bibinfo
  {journal} {AIP Advances}\ }\textbf {\bibinfo {volume} {6}},\ \bibinfo {pages}
  {115021} (\bibinfo {year} {2016})}\BibitemShut {NoStop}%
\bibitem [{\citenamefont {Ke}\ \emph {et~al.}(2019)\citenamefont {Ke},
  \citenamefont {Moehle}, \citenamefont {Vries}, \citenamefont {Thomas},
  \citenamefont {Metti}, \citenamefont {Guinn}, \citenamefont {Kallaher},
  \citenamefont {Lodari}, \citenamefont {Scappucci}, \citenamefont {Wang},
  \citenamefont {Diaz}, \citenamefont {Gardner}, \citenamefont {Manfra},\ and\
  \citenamefont {Goswami}}]{10.1038/s41467-019-11742-4}%
  \BibitemOpen
  \bibfield  {author} {\bibinfo {author} {\bibfnamefont {C.~T.}\ \bibnamefont
  {Ke}}, \bibinfo {author} {\bibfnamefont {C.~M.}\ \bibnamefont {Moehle}},
  \bibinfo {author} {\bibfnamefont {F.~K.~d.}\ \bibnamefont {Vries}}, \bibinfo
  {author} {\bibfnamefont {C.}~\bibnamefont {Thomas}}, \bibinfo {author}
  {\bibfnamefont {S.}~\bibnamefont {Metti}}, \bibinfo {author} {\bibfnamefont
  {C.~R.}\ \bibnamefont {Guinn}}, \bibinfo {author} {\bibfnamefont
  {R.}~\bibnamefont {Kallaher}}, \bibinfo {author} {\bibfnamefont
  {M.}~\bibnamefont {Lodari}}, \bibinfo {author} {\bibfnamefont
  {G.}~\bibnamefont {Scappucci}}, \bibinfo {author} {\bibfnamefont
  {T.}~\bibnamefont {Wang}}, \bibinfo {author} {\bibfnamefont {R.~E.}\
  \bibnamefont {Diaz}}, \bibinfo {author} {\bibfnamefont {G.~C.}\ \bibnamefont
  {Gardner}}, \bibinfo {author} {\bibfnamefont {M.~J.}\ \bibnamefont
  {Manfra}},\ and\ \bibinfo {author} {\bibfnamefont {S.}~\bibnamefont
  {Goswami}},\ }\bibfield  {title} {\bibinfo {title} {{Ballistic
  superconductivity and tunable pi–junctions in InSb quantum wells}},\ }\href
  {https://doi.org/10.1038/s41467-019-11742-4} {\bibfield  {journal} {\bibinfo
  {journal} {Nature Communications}\ }\textbf {\bibinfo {volume} {10}},\
  \bibinfo {pages} {3764} (\bibinfo {year} {2019})}\BibitemShut {NoStop}%
\bibitem [{\citenamefont {Salimian}\ \emph {et~al.}(2021)\citenamefont
  {Salimian}, \citenamefont {Carrega}, \citenamefont {Verma}, \citenamefont
  {Zannier}, \citenamefont {Nowak}, \citenamefont {Beltram}, \citenamefont
  {Sorba},\ and\ \citenamefont {Heun}}]{10.1063/5.0071218}%
  \BibitemOpen
  \bibfield  {author} {\bibinfo {author} {\bibfnamefont {S.}~\bibnamefont
  {Salimian}}, \bibinfo {author} {\bibfnamefont {M.}~\bibnamefont {Carrega}},
  \bibinfo {author} {\bibfnamefont {I.}~\bibnamefont {Verma}}, \bibinfo
  {author} {\bibfnamefont {V.}~\bibnamefont {Zannier}}, \bibinfo {author}
  {\bibfnamefont {M.~P.}\ \bibnamefont {Nowak}}, \bibinfo {author}
  {\bibfnamefont {F.}~\bibnamefont {Beltram}}, \bibinfo {author} {\bibfnamefont
  {L.}~\bibnamefont {Sorba}},\ and\ \bibinfo {author} {\bibfnamefont
  {S.}~\bibnamefont {Heun}},\ }\bibfield  {title} {\bibinfo {title}
  {{Gate-controlled supercurrent in ballistic InSb nanoflag Josephson
  junctions}},\ }\href {https://doi.org/10.1063/5.0071218} {\bibfield
  {journal} {\bibinfo  {journal} {Applied Physics Letters}\ }\textbf {\bibinfo
  {volume} {119}},\ \bibinfo {pages} {214004} (\bibinfo {year}
  {2021})}\BibitemShut {NoStop}%
\bibitem [{\citenamefont {Houver}\ \emph {et~al.}(2019)\citenamefont {Houver},
  \citenamefont {Huber}, \citenamefont {Savoini}, \citenamefont {Abreu},\ and\
  \citenamefont {Johnson}}]{10.1364/OE.27.010854}%
  \BibitemOpen
  \bibfield  {author} {\bibinfo {author} {\bibfnamefont {S.}~\bibnamefont
  {Houver}}, \bibinfo {author} {\bibfnamefont {L.}~\bibnamefont {Huber}},
  \bibinfo {author} {\bibfnamefont {M.}~\bibnamefont {Savoini}}, \bibinfo
  {author} {\bibfnamefont {E.}~\bibnamefont {Abreu}},\ and\ \bibinfo {author}
  {\bibfnamefont {S.~L.}\ \bibnamefont {Johnson}},\ }\bibfield  {title}
  {\bibinfo {title} {{2D THz spectroscopic investigation of ballistic
  conduction-band electron dynamics in InSb}},\ }\href
  {https://doi.org/10.1364/oe.27.010854} {\bibfield  {journal} {\bibinfo
  {journal} {Optics Express}\ }\textbf {\bibinfo {volume} {27}},\ \bibinfo
  {pages} {10854} (\bibinfo {year} {2019})}\BibitemShut {NoStop}%
\bibitem [{\citenamefont {Kress}\ \emph {et~al.}(2004)\citenamefont {Kress},
  \citenamefont {Löffler}, \citenamefont {Eden}, \citenamefont {Thomson},\
  and\ \citenamefont {Roskos}}]{10.1364/OL.29.001120}%
  \BibitemOpen
  \bibfield  {author} {\bibinfo {author} {\bibfnamefont {M.}~\bibnamefont
  {Kress}}, \bibinfo {author} {\bibfnamefont {T.}~\bibnamefont {Löffler}},
  \bibinfo {author} {\bibfnamefont {S.}~\bibnamefont {Eden}}, \bibinfo {author}
  {\bibfnamefont {M.}~\bibnamefont {Thomson}},\ and\ \bibinfo {author}
  {\bibfnamefont {H.~G.}\ \bibnamefont {Roskos}},\ }\bibfield  {title}
  {\bibinfo {title} {{Terahertz-pulse generation by photoionization of air with
  laser pulses composed of both fundamental and second-harmonic waves}},\
  }\href {https://doi.org/10.1364/ol.29.001120} {\bibfield  {journal} {\bibinfo
   {journal} {Optics Letters}\ }\textbf {\bibinfo {volume} {29}},\ \bibinfo
  {pages} {1120} (\bibinfo {year} {2004})}\BibitemShut {NoStop}%
\bibitem [{\citenamefont {Yang}\ \emph {et~al.}(2007)\citenamefont {Yang},
  \citenamefont {Mutter}, \citenamefont {Stillhart}, \citenamefont {Ruiz},
  \citenamefont {Aravazhi}, \citenamefont {Jazbinsek}, \citenamefont
  {Schneider}, \citenamefont {Gramlich},\ and\ \citenamefont
  {Günter}}]{10.1002/adfm.200601117}%
  \BibitemOpen
  \bibfield  {author} {\bibinfo {author} {\bibfnamefont {Z.}~\bibnamefont
  {Yang}}, \bibinfo {author} {\bibfnamefont {L.}~\bibnamefont {Mutter}},
  \bibinfo {author} {\bibfnamefont {M.}~\bibnamefont {Stillhart}}, \bibinfo
  {author} {\bibfnamefont {B.}~\bibnamefont {Ruiz}}, \bibinfo {author}
  {\bibfnamefont {S.}~\bibnamefont {Aravazhi}}, \bibinfo {author}
  {\bibfnamefont {M.}~\bibnamefont {Jazbinsek}}, \bibinfo {author}
  {\bibfnamefont {A.}~\bibnamefont {Schneider}}, \bibinfo {author}
  {\bibfnamefont {V.}~\bibnamefont {Gramlich}},\ and\ \bibinfo {author}
  {\bibfnamefont {P.}~\bibnamefont {Günter}},\ }\bibfield  {title} {\bibinfo
  {title} {{Large‐Size Bulk and Thin‐Film Stilbazolium‐Salt Single
  Crystals for Nonlinear Optics and THz Generation}},\ }\href
  {https://doi.org/10.1002/adfm.200601117} {\bibfield  {journal} {\bibinfo
  {journal} {Advanced Functional Materials}\ }\textbf {\bibinfo {volume}
  {17}},\ \bibinfo {pages} {2018} (\bibinfo {year} {2007})}\BibitemShut
  {NoStop}%
\bibitem [{\citenamefont {Dai}\ \emph {et~al.}(2006)\citenamefont {Dai},
  \citenamefont {Xie},\ and\ \citenamefont
  {Zhang}}]{10.1103/PhysRevLett.97.103903}%
  \BibitemOpen
  \bibfield  {author} {\bibinfo {author} {\bibfnamefont {J.}~\bibnamefont
  {Dai}}, \bibinfo {author} {\bibfnamefont {X.}~\bibnamefont {Xie}},\ and\
  \bibinfo {author} {\bibfnamefont {X.-C.}\ \bibnamefont {Zhang}},\ }\bibfield
  {title} {\bibinfo {title} {{Detection of Broadband Terahertz Waves with a
  Laser-Induced Plasma in Gases}},\ }\href
  {https://doi.org/10.1103/physrevlett.97.103903} {\bibfield  {journal}
  {\bibinfo  {journal} {Physical Review Letters}\ }\textbf {\bibinfo {volume}
  {97}},\ \bibinfo {pages} {103903} (\bibinfo {year} {2006})}\BibitemShut
  {NoStop}%
\bibitem [{\citenamefont {Yee}(1966)}]{10.1109/TAP.1966.1138693}%
  \BibitemOpen
  \bibfield  {author} {\bibinfo {author} {\bibfnamefont {K.}~\bibnamefont
  {Yee}},\ }\bibfield  {title} {\bibinfo {title} {{Numerical solution of
  initial boundary value problems involving maxwell's equations in isotropic
  media}},\ }\href {https://doi.org/10.1109/tap.1966.1138693} {\bibfield
  {journal} {\bibinfo  {journal} {IEEE Transactions on Antennas and
  Propagation}\ }\textbf {\bibinfo {volume} {14}},\ \bibinfo {pages} {302}
  (\bibinfo {year} {1966})}\BibitemShut {NoStop}%
\bibitem [{\citenamefont {Yu}\ \emph {et~al.}(2017)\citenamefont {Yu},
  \citenamefont {Heffernan},\ and\ \citenamefont
  {Talbayev}}]{10.1103/PhysRevB.95.125201}%
  \BibitemOpen
  \bibfield  {author} {\bibinfo {author} {\bibfnamefont {S.}~\bibnamefont
  {Yu}}, \bibinfo {author} {\bibfnamefont {K.~H.}\ \bibnamefont {Heffernan}},\
  and\ \bibinfo {author} {\bibfnamefont {D.}~\bibnamefont {Talbayev}},\
  }\bibfield  {title} {\bibinfo {title} {{Beyond the effective mass
  approximation: A predictive theory of the nonlinear optical response of
  conduction electrons}},\ }\href {https://doi.org/10.1103/physrevb.95.125201}
  {\bibfield  {journal} {\bibinfo  {journal} {Physical Review B}\ }\textbf
  {\bibinfo {volume} {95}},\ \bibinfo {pages} {125201} (\bibinfo {year}
  {2017})}\BibitemShut {NoStop}%
\bibitem [{\citenamefont {Devreese}\ \emph {et~al.}(1982)\citenamefont
  {Devreese}, \citenamefont {Welzenis},\ and\ \citenamefont
  {Evrard}}]{10.1007/BF00617768}%
  \BibitemOpen
  \bibfield  {author} {\bibinfo {author} {\bibfnamefont {J.~T.}\ \bibnamefont
  {Devreese}}, \bibinfo {author} {\bibfnamefont {R.~G.~v.}\ \bibnamefont
  {Welzenis}},\ and\ \bibinfo {author} {\bibfnamefont {R.~P.}\ \bibnamefont
  {Evrard}},\ }\bibfield  {title} {\bibinfo {title} {{Impact ionisation
  probability in InSb}},\ }\href {https://doi.org/10.1007/bf00617768}
  {\bibfield  {journal} {\bibinfo  {journal} {Applied Physics A}\ }\textbf
  {\bibinfo {volume} {29}},\ \bibinfo {pages} {125} (\bibinfo {year}
  {1982})}\BibitemShut {NoStop}%
\bibitem [{\citenamefont {Ašmontas}\ \emph {et~al.}(2013)\citenamefont
  {Ašmontas}, \citenamefont {Raguotis},\ and\ \citenamefont
  {Bumelienė}}]{10.1088/0268-1242/28/2/025019}%
  \BibitemOpen
  \bibfield  {author} {\bibinfo {author} {\bibfnamefont {S.}~\bibnamefont
  {Ašmontas}}, \bibinfo {author} {\bibfnamefont {R.}~\bibnamefont
  {Raguotis}},\ and\ \bibinfo {author} {\bibfnamefont {S.}~\bibnamefont
  {Bumelienė}},\ }\bibfield  {title} {\bibinfo {title} {{Monte Carlo
  calculations of the electron impact ionization in n-type InSb crystal}},\
  }\href {https://doi.org/10.1088/0268-1242/28/2/025019} {\bibfield  {journal}
  {\bibinfo  {journal} {Semiconductor Science and Technology}\ }\textbf
  {\bibinfo {volume} {28}},\ \bibinfo {pages} {025019} (\bibinfo {year}
  {2013})}\BibitemShut {NoStop}%
\bibitem [{\citenamefont {Ašmontas}\ \emph {et~al.}(2020)\citenamefont
  {Ašmontas}, \citenamefont {Bumelienė}, \citenamefont {Gradauskas},
  \citenamefont {Raguotis},\ and\ \citenamefont
  {Sužiedėlis}}]{10.1038/s41598-020-67541-1}%
  \BibitemOpen
  \bibfield  {author} {\bibinfo {author} {\bibfnamefont {S.}~\bibnamefont
  {Ašmontas}}, \bibinfo {author} {\bibfnamefont {S.}~\bibnamefont
  {Bumelienė}}, \bibinfo {author} {\bibfnamefont {J.}~\bibnamefont
  {Gradauskas}}, \bibinfo {author} {\bibfnamefont {R.}~\bibnamefont
  {Raguotis}},\ and\ \bibinfo {author} {\bibfnamefont {A.}~\bibnamefont
  {Sužiedėlis}},\ }\bibfield  {title} {\bibinfo {title} {{Impact ionization
  and intervalley electron scattering in InSb and InAs induced by a single
  terahertz pulse}},\ }\href {https://doi.org/10.1038/s41598-020-67541-1}
  {\bibfield  {journal} {\bibinfo  {journal} {Scientific Reports}\ }\textbf
  {\bibinfo {volume} {10}},\ \bibinfo {pages} {10580} (\bibinfo {year}
  {2020})}\BibitemShut {NoStop}%
\bibitem [{\citenamefont {Kim}\ \emph {et~al.}(2009)\citenamefont {Kim},
  \citenamefont {Hummer},\ and\ \citenamefont
  {Kresse}}]{10.1103/PhysRevB.80.035203}%
  \BibitemOpen
  \bibfield  {author} {\bibinfo {author} {\bibfnamefont {Y.-S.}\ \bibnamefont
  {Kim}}, \bibinfo {author} {\bibfnamefont {K.}~\bibnamefont {Hummer}},\ and\
  \bibinfo {author} {\bibfnamefont {G.}~\bibnamefont {Kresse}},\ }\bibfield
  {title} {\bibinfo {title} {{Accurate band structures and effective masses for
  InP, InAs, and InSb using hybrid functionals}},\ }\href
  {https://doi.org/10.1103/physrevb.80.035203} {\bibfield  {journal} {\bibinfo
  {journal} {Physical Review B}\ }\textbf {\bibinfo {volume} {80}},\ \bibinfo
  {pages} {035203} (\bibinfo {year} {2009})}\BibitemShut {NoStop}%
\bibitem [{\citenamefont {Kane}(1957)}]{10.1016/0022-3697(57)90013-6}%
  \BibitemOpen
  \bibfield  {author} {\bibinfo {author} {\bibfnamefont {E.~O.}\ \bibnamefont
  {Kane}},\ }\bibfield  {title} {\bibinfo {title} {{Band structure of indium
  antimonide}},\ }\href {https://doi.org/10.1016/0022-3697(57)90013-6}
  {\bibfield  {journal} {\bibinfo  {journal} {Journal of Physics and Chemistry
  of Solids}\ }\textbf {\bibinfo {volume} {1}},\ \bibinfo {pages} {249}
  (\bibinfo {year} {1957})}\BibitemShut {NoStop}%
\bibitem [{\citenamefont {Zawadzki}(1974)}]{10.1080/00018737400101371}%
  \BibitemOpen
  \bibfield  {author} {\bibinfo {author} {\bibfnamefont {W.}~\bibnamefont
  {Zawadzki}},\ }\bibfield  {title} {\bibinfo {title} {{Electron transport
  phenomena in small-gap semiconductors}},\ }\href
  {https://doi.org/10.1080/00018737400101371} {\bibfield  {journal} {\bibinfo
  {journal} {Advances in Physics}\ }\textbf {\bibinfo {volume} {23}},\ \bibinfo
  {pages} {435} (\bibinfo {year} {1974})}\BibitemShut {NoStop}%
\bibitem [{\citenamefont {Hirori}\ \emph {et~al.}(2011)\citenamefont {Hirori},
  \citenamefont {Shinokita}, \citenamefont {Shirai}, \citenamefont {Tani},
  \citenamefont {Kadoya},\ and\ \citenamefont {Tanaka}}]{10.1038/ncomms1598}%
  \BibitemOpen
  \bibfield  {author} {\bibinfo {author} {\bibfnamefont {H.}~\bibnamefont
  {Hirori}}, \bibinfo {author} {\bibfnamefont {K.}~\bibnamefont {Shinokita}},
  \bibinfo {author} {\bibfnamefont {M.}~\bibnamefont {Shirai}}, \bibinfo
  {author} {\bibfnamefont {S.}~\bibnamefont {Tani}}, \bibinfo {author}
  {\bibfnamefont {Y.}~\bibnamefont {Kadoya}},\ and\ \bibinfo {author}
  {\bibfnamefont {K.}~\bibnamefont {Tanaka}},\ }\bibfield  {title} {\bibinfo
  {title} {{Extraordinary carrier multiplication gated by a picosecond electric
  field pulse}},\ }\href {https://doi.org/10.1038/ncomms1598} {\bibfield
  {journal} {\bibinfo  {journal} {Nature Communications}\ }\textbf {\bibinfo
  {volume} {2}},\ \bibinfo {pages} {594} (\bibinfo {year} {2011})}\BibitemShut
  {NoStop}%
\bibitem [{\citenamefont {Tanimura}\ \emph {et~al.}(2015)\citenamefont
  {Tanimura}, \citenamefont {Kanasaki},\ and\ \citenamefont
  {Tanimura}}]{10.1103/PhysRevB.91.045201}%
  \BibitemOpen
  \bibfield  {author} {\bibinfo {author} {\bibfnamefont {H.}~\bibnamefont
  {Tanimura}}, \bibinfo {author} {\bibfnamefont {J.}~\bibnamefont {Kanasaki}},\
  and\ \bibinfo {author} {\bibfnamefont {K.}~\bibnamefont {Tanimura}},\
  }\bibfield  {title} {\bibinfo {title} {{Ultrafast scattering processes of hot
  electrons in InSb studied by time- and angle-resolved photoemission
  spectroscopy}},\ }\href {https://doi.org/10.1103/physrevb.91.045201}
  {\bibfield  {journal} {\bibinfo  {journal} {Physical Review B}\ }\textbf
  {\bibinfo {volume} {91}},\ \bibinfo {pages} {045201} (\bibinfo {year}
  {2015})}\BibitemShut {NoStop}%
\end{thebibliography}%

\end{document}